\documentclass[dvips,12pt] {article}
\usepackage{epsfig}
\usepackage{a4}
\newcommand{\be}{\begin{equation}}
\newcommand{\ee}{\end{equation}}
\newcommand{\bea}{\begin{eqnarray}}
\newcommand{\eea}{\end{eqnarray}}
\newcommand{\ep}{\varepsilon}

\newcommand{\nn}{\nonumber}
\newcommand{\half}{{\textstyle{1\over2}}}
\newcommand{\halfn}{{\textstyle{n\over2}}}

\def\Month{\ifcase\month\or
January\or February\or March\or April\or May\or June\or
July\or August\or September\or October\or November\or December\fi}

\begin{document}
 \thispagestyle{empty}
 \begin{flushright}
 {CERN-TH/98-211} \\[2mm]
 {MZ-TH/98-22} \\[3mm]
 {hep-ph/9806522} 
\end{flushright}
 \vspace*{2.0cm}
 \begin{center}
 {\bf \Large
 On-shell \ two-loop \ three-gluon \ vertex }
 \end{center}
 \vspace{1cm}
 \begin{center}
 A.~I.~Davydychev$^{a,b,}$\footnote{davyd@thep.physik.uni-mainz.de} 
 \ \ and \ \
 P.~Osland$^{c,d,e,}$\footnote{Per.Osland@fi.uib.no} 
\\
 \vspace{1cm}
$^{a}${\em
 Department of Physics,
 University of Mainz, \\
 Staudingerweg 7,
 D-55099 Mainz, Germany}
\\
\vspace{.3cm}
$^{b}${\em
 Institute for Nuclear Physics,
 Moscow State University, \\
 119899, Moscow, Russia}
\\
\vspace{.3cm}
$^{c}${\em
 Department of Physics, University of Bergen, \\
      All\'{e}gaten 55, N-5007 Bergen, Norway}
\\
\vspace{.3cm}
$^{d}${\em
 Deutsches Elektronen-Synchrotron DESY,
 D-22603 Hamburg, Germany}
\\
\vspace{.3cm}
$^{e}${\em
Theoretical Physics Division, CERN, 
CH-1211 Geneva 23, Switzerland }
\end{center}
 \hspace{3in}
 \begin{abstract}
The two-loop three-gluon vertex is calculated in an arbitrary covariant
gauge, in the limit when two of the gluons are on the mass shell.
The corresponding two-loop results for the ghost-gluon vertex are also 
obtained. It is shown that the results are consistent with 
the Ward--Slavnov--Taylor identities. 
 \end{abstract}

\vfill
\begin{flushleft}
 {CERN-TH/98-211} \\[2mm]
 {MZ-TH/98-22} \\[3mm]
 {June 1998}
\end{flushleft}
\newpage

\setcounter{footnote}{0}  


\section{Introduction}
\setcounter{equation}{0}

Calculation of radiative corrections to certain jet processes
is becoming extremely important (see, for instance, the reviews 
\cite{reviews,Catani-rev} and references therein).
In particular, the two-loop three-gluon vertex with two
external gluons on shell (i.e.\ when two external momenta
squared vanish) is a part of the 
next-to-next-to-leading order (NNLO) contributions 
to $gg\rightarrow gg$ and $q\overline{q}\rightarrow gg$
processes (see, for instance, Fig.~1).
The next-to-leading order (NLO) contributions to these processes
were considered in ref.~\cite{ElSex}\footnote{For a complete list
of NLO contributions, see Figs.~6 and 8 of \cite{ElSex}.}
(see also in refs.~\cite{jets}).  
For a complete calculation of all relevant NNLO contributions,
one also needs to calculate the two-loop four-point functions 
(see, for instance, in \cite{Uss,BRY}).

Studying the three-gluon vertex (and other QCD vertices) 
in this on-shell limit is also important, in order to understand
the structure of infrared (on-shell) singularities, to illustrate how
the corresponding Ward--Slavnov--Taylor (WST) identities work 
in this divergent case
and to develop calculational techniques that can then be 
generalized to deal with more complicated (e.g.\ off-shell)
configurations. Performing the calculations in an arbitrary covariant 
gauge is useful, in order to exploit the consequences
of gauge invariance (the WST identities) at all stages. 
In the calculation of physical quantities, the independence
of the gauge parameter usually serves as an important check.

The one-loop QCD vertices have been known for quite some time.
The one-loop result for the three-gluon vertex, for off-shell gluons
(with $p_1^2=p_2^2=p_3^3$) and in an arbitrary covariant gauge, 
was presented in \cite{CG,PT}. The general off-shell case,
but restricted to the Feynman gauge, was considered in \cite{BC2}.
Various on-shell results have also been given: in \cite{BF},
restricted to the infrared-singular parts only
(in an arbitrary covariant gauge),
and in \cite{NPS}, with the finite parts for the case of
two gluons being on-shell (in the Feynman gauge).
The most general results, valid for arbitrary values
of the space-time dimension and the covariant-gauge
parameter, have been presented in an earlier paper \cite{DOT1},
where we have also collected the results for all on-shell
limits of interest. 
Some results for the one-loop quark--gluon vertex (or its
Abelian part, which is related to the QED vertex) can be
found in \cite{qqg}.

At the two-loop level, the QCD vertices were mainly studied in the
zero-momentum limit \cite{BL,DOT2}, i.e.\ when one of the external
momenta is zero. This limit is useful for studying the 
renormalization properties of QCD, since (for the considered
vertices) it does not bring in any infrared (on-shell) singularities.
In \cite{BL}, the renormalized results in the Feynman gauge
were presented. In \cite{DOT2}, the unrenormalized and renormalized
results for the three-gluon and ghost-gluon vertices have been
obtained in an arbitrary covariant gauge, and the corresponding 
differential WST identity in QCD was analysed in detail.
The relevant techniques for on-shell two-loop calculations
have been studied in refs.~\cite{Gons,Neerven,KL}.
In ref.~\cite{Gons}, the two-loop electromagnetic quark form
factor in massless QCD was considered. Corrected results for
this form factor were later presented in \cite{MvN,KL2}.
In ref.~\cite{Neerven}, the two-loop scalar form factor was
calculated. We also note that
the structure and factorization properties of infrared singularities 
of two-loop order QCD amplitudes were recently discussed 
in ref.~\cite{Catani}.

In the present paper, we discuss an algorithm to calculate 
two-loop three-point diagrams with two external legs on shell.
Then, we present the on-shell results for the three-gluon vertex
and the ghost-gluon vertex, in an arbitrary covariant gauge,
keeping the finite parts of the expansion in the dimensional
regularization \cite{dimreg} parameter $\ep$.
We consider the relevant WST identities in the on-shell limit
and confirm that the results obtained are consistent
with these identities. 

\section{Preliminaries and WST identities}
\setcounter{equation}{0}

The three-gluon vertex is defined as (see Fig.~2)
\begin{equation}
\label{ggg}
\Gamma_{\mu_1 \mu_2 \mu_3}^{a_1 a_2 a_3}(p_1, p_2, p_3)
\equiv  - \mbox{i} \; g \;
f^{a_1 a_2 a_3} \; \Gamma_{\mu_1 \mu_2 \mu_3}(p_1, p_2, p_3) .
\end{equation}
Here, the $f^{a_1 a_2 a_3}$
are the totally antisymmetric colour structures corresponding  
to the adjoint representation of the gauge group
(for example, $\mbox{SU}(N)$ or any other semi-simple 
gauge group).
Other colour structures do not appear in the perturbative 
calculation of QCD three-point vertices, at least at the one- and 
two-loop levels.
To regulate the ultraviolet and infrared (on-shell) divergences 
occurring at the one- and two-loop levels,
we shall use dimensional regularization \cite{dimreg},
with the space-time dimension $n=4-2\ep$.

Since the $f^{a_1 a_2 a_3}$ are antisymmetric,
$\Gamma_{\mu_1 \mu_2 \mu_3}(p_1, p_2, p_3)$ must also be
{\em antisymmetric}
under any interchange of a pair of gluon momenta and the
corresponding Lorentz indices.
Therefore, in the limit of interest
($p_1^2=p_2^2=0, \; p_3^2\equiv p^2$) it can
be presented as
\bea
\label{ggg-U}
\left. \Gamma_{\mu_1 \mu_2 \mu_3}(p_1, p_2, p_3)\right|_{
p_1^2=p_2^2=0, \; p_3^2\equiv p^2} 
= g_{\mu_1 \mu_2} \; (p_1 - p_2)_{\mu_3} \; U_1(p^2)
\hspace{45mm}
\nn \\
+ \left[ g_{\mu_1 \mu_3} p_{1 \mu_2}
       - g_{\mu_2 \mu_3} p_{2 \mu_1} \right] \; U_2(p^2)
+ \left[ g_{\mu_1 \mu_3} p_{2 \mu_2}
       - g_{\mu_2 \mu_3} p_{1 \mu_1} \right] \; U_3(p^2)
\hspace{23mm}
\nn \\
+ p_{1 \mu_1} p_{2 \mu_2} (p_1-p_2)_{\mu_3} \; U_4(p^2)
+ p_{2 \mu_1} p_{1 \mu_2} (p_1-p_2)_{\mu_3} \; U_5(p^2)
\hspace{36mm}
\nn \\
+ \left[ p_{1 \mu_1} p_{1 \mu_2} p_{1 \mu_3}
       - p_{2 \mu_1} p_{2 \mu_2} p_{2 \mu_3} \right] \; U_6(p^2)
+ \left[ p_{1 \mu_1} p_{1 \mu_2} p_{2 \mu_3}
       - p_{2 \mu_1} p_{2 \mu_2} p_{1 \mu_3} \right] \; U_7(p^2) .
\eea
This decomposition is similar
to eq.~(29) of ref.~\cite{BF}\footnote{For details,
see section~4C and appendix~F of ref.~\cite{DOT1}.}.
All terms are explicitly antisymmetric with respect to 
$(p_1, \mu_1) \leftrightarrow (p_2, \mu_2)$.
At the lowest, ``zero-loop'' order,
\be
\label{U0}
U_1^{(0)}=1, \hspace{5mm}
U_2^{(0)}=-2, \hspace{5mm}
U_3^{(0)}=-1, \hspace{5mm}
U_4^{(0)}=U_5^{(0)}=U_6^{(0)}=U_7^{(0)}=0,
\ee
so we get the well-known tensor structure
\be
g_{\mu_1 \mu_2} (p_1 - p_2)_{\mu_3}
      + g_{\mu_2 \mu_3} (p_2 - p_3)_{\mu_1}
      + g_{\mu_3 \mu_1} (p_3 - p_1)_{\mu_2}.
\ee

The lowest-order gluon propagator is\footnote{Here
and henceforth, a causal prescription is understood,
$1/p^2 \rightarrow 1/(p^2 +\mbox{i}0)$.}
\begin{equation}
\label{gl_prop}
\delta^{a_1 a_2} \; \frac{1}{p^2}
\left( g_{\mu_1 \mu_2} - \xi \; \frac{p_{\mu_1} p_{\mu_2}}{p^2}
\right) ,
\end{equation}
where $\xi\equiv1-\alpha$ is the gauge parameter corresponding
to a general covariant gauge,
defined in such a way that $\xi=0$ ($\alpha=1$) is the Feynman gauge.
The gluon polarization operator is defined as
\begin{equation}
\label{gl_po}
\Pi_{\mu_1 \mu_2}^{a_1 a_2}(p)
\equiv - \delta^{a_1 a_2}
\left( p^2 g_{\mu_1 \mu_2} - {p}_{\mu_1}{p}_{\mu_2} \right) J(p^2),
\end{equation}
while the ghost self-energy is
\begin{equation}
\label{gh_se}
\widetilde{\Pi}^{a_1 a_2}(p^2) = \delta^{a_1 a_2} \; p^2 \;
\left[G(p^2)\right]^{-1} .
\end{equation}  
In the lowest-order approximation $J^{(0)}=G^{(0)}=1$. 

The ghost-gluon vertex can be represented as   
\be
\label{ghg}
\widetilde{\Gamma}_{\mu_3}^{a_1 a_2 a_3}(p_1, p_2; p_3)
\equiv -\mbox{i} g \; f^{a_1 a_2 a_3} \;
{p_1}^{\mu} \; \widetilde{\Gamma}_{\mu \mu_3}(p_1, p_2; p_3) ,
\ee
where $p_1$ is the out-ghost momentum, $p_2$ is the in-ghost momentum,
$p_3$ and $\mu_3$ are the momentum and the Lorentz index of the gluon
(all momenta are ingoing).
For $\widetilde{\Gamma}_{\mu \mu_3}$, we use following 
decomposition (see \cite{BC2}\footnote{Adopting the notation 
used in ref.~\cite{BC2},
we have written in previous papers \cite{DOT1,DOT2} the arguments
of these ghost-gluon scalar functions $a$, $b$, $c$, $d$ and $e$ as 
momenta (vectors).
In the present paper, it is more convenient to 
write the arguments as momenta {\em squared}.
The {\em order} (3,2,1) of the arguments on the r.h.s.\ of 
eq.~(\ref{BC-ghg}) corresponds to ref.~\cite{BC2}
and has not been changed.}):
\begin{eqnarray}
\label{BC-ghg}
\widetilde{\Gamma}_{\mu \mu_3}(p_1,p_2;p_3)
= g_{\mu \mu_3} a(p_3^2,p_2^2,p_1^2)
- {p_3}_{\mu} {p_2}_{\mu_3} b(p_3^2,p_2^2,p_1^2)
+ {p_1}_{\mu} {p_3}_{\mu_3} c(p_3^2,p_2^2,p_1^2)
\nonumber \\
+ {p_3}_{\mu} {p_1}_{\mu_3} d(p_3^2,p_2^2,p_1^2) 
+ {p_1}_{\mu} {p_1}_{\mu_3} e(p_3^2,p_2^2,p_1^2) .
\end{eqnarray}
At the ``zero-loop'' level,
all the scalar functions involved in  (\ref{BC-ghg})
vanish at this order, except one, $a^{(0)}=1$.

Whenever possible, we adopt the notation used in our previous papers
\cite{DOT1,DOT2}. In particular,
for a quantity $X$ (e.g.\ any of the scalar functions contributing 
to the propagators or the vertices), we denote the
zero-loop order contribution as $X^{(0)}$ 
(cf.\ eq.~(\ref{U0})),
the one-loop order contribution as
$X^{(1)}$, and the two-loop order contribution as $X^{(2)}$.
In this paper, as a rule,
\begin{equation}
X^{(L)}=X^{(L,\xi)} + X^{(L,q)},
\end{equation}
where $X^{(L,\xi)}$ denotes the contribution
of gluon and ghost
loops in a general covariant gauge (\ref{gl_prop}) (in particular,
$X^{(L,0)}$ corresponds to the Feynman gauge, $\xi=0$),  
while $X^{(L,q)}$ represents the contribution of the
quark loops.

In general, the WST identity \cite{WST} 
for the three-gluon vertex reads (see e.g.\ in \cite{MarPag}):
\bea
\label{WST3}
p_3^{\mu_3} \; \Gamma_{\mu_1 \mu_2 \mu_3}(p_1,p_2,p_3)
= -J(p_1^2) \; G(p_3^2) \; 
\left[ g_{\mu_1}^{\;\;\; \mu_3} p_1^2 - p_{1 \mu_1} p_1^{\mu_3} \right]
\; \widetilde{\Gamma}_{\mu_3 \mu_2}(p_1,p_3; p_2)
\nn \\
+ J(p_2^2) \; G(p_3^2) \;
\left[ g_{\mu_2}^{\;\;\; \mu_3} p_2^2 - p_{2 \mu_2} p_2^{\mu_3} \right]
\; \widetilde{\Gamma}_{\mu_3 \mu_1}(p_2,p_3; p_1) .
\eea
It is easy to see that the $c$ and $e$ functions from
the ghost-gluon vertex (\ref{BC-ghg}) do not
contribute to the WST identity (\ref{WST3}).
In the on-shell case, some of the momenta squared vanish.
Note that (in the case of massless quarks) $J(0)=G(0)=1$. 
In eq.~(\ref{WST3}) we can also consider permutations 
of the indices 1, 2, 3 corresponding
to the contractions of the three-gluon vertex with $p_1^{\mu_1}$
or $p_2^{\mu_2}$. To get all relations between the scalar
functions in the case of interest ($p_1^2=p_2^2=0, \; p_3^2\equiv p^2$),
it is sufficient to consider the contractions with $p_3^{\mu_3}$ and
$p_1^{\mu_1}$.

The WST identity (\ref{WST3}) (contraction with $p_3^{\mu_3}$) yields
just one condition on the scalar functions:
\bea
\label{WST3-1}
U_2(p^2) - U_3(p^2) + \half p^2 U_6(p^2) + \half p^2 U_7(p^2)
\hspace{60mm}
\nn \\
= - J(0) \; G(p^2) 
\left[ a(0,p^2,0)+\half p^2 b(0,p^2,0)+\half p^2 d(0,p^2,0) \right].
\eea

Considering contraction with $p_1^{\mu_1}$, we get
four relations:
\be
\label{WST1-1}
U_1(p^2)+\half p^2 U_5(p^2) = G(0)\; J(p^2)
\left[ a(0,0,p^2) + \half p^2 b(0,0,p^2) + \half p^2 d(0,0,p^2) \right] ,
\ee
\be
\label{WST1-2}
U_2(p^2) = -2\; G(0) \; J(p^2) \; a(0,0,p^2) ,
\ee
\be
\label{WST1-3}
U_3(p^2) \!-\! \half p^2 U_7(p^2) 
= G(0) J(0)\; \half p^2 b(p^2,0,0)
- G(0) J(p^2) \left[ a(0,0,p^2) \!-\! \half p^2 d(0,0,p^2) \right] ,
\ee
\bea
\label{WST1-4}
\half p^2 U_6(p^2) = - G(0)\; J(0)\;  
\left[ a(p^2,0,0) - \half p^2 d(p^2,0,0)  \right]
\nn \\
+ G(0)\; J(p^2) 
\left[ a(0,0,p^2) + \half p^2 d(0,0,p^2) \right] .
\eea

Note that the l.h.s.\ of eq.~(\ref{WST3-1}) can be constructed from
the l.h.s.'s of eqs.~(\ref{WST1-2})--(\ref{WST1-4}).
This gives a condition on the scalar functions from the
ghost-gluon vertex:
\bea
\label{WST-ghg}
G(p^2) \; \left[ a(0,p^2,0) + \half p^2 b(0,p^2,0) 
                 + \half p^2 d(0,p^2,0) \right]
\hspace{40mm}
\nn \\
=
G(0) \; \left[ a(p^2,0,0) + \half p^2 b(p^2,0,0)
                       - \half p^2 d(p^2,0,0) \right] .
\eea
Therefore, the conditions (\ref{WST1-1})--(\ref{WST-ghg}) may be
considered as a set of independent corollaries of the WST identities
(\ref{WST3}). 

At the one-loop order, the diagrams contributing to the three-gluon
vertex, two-point functions and the ghost-gluon vertex are
shown in Figs.~1, 2 and 3 of ref.~\cite{DOT1}.
The one-loop expressions for the $U_i$ functions are presented
in Appendix~F of ref.~\cite{DOT1}\footnote{The divergent parts
were presented earlier in ref.~\cite{BF}, whereas the Feynman-gauge
results, including the finite terms in $\ep$, are available
in \cite{NPS}.}, for arbitrary values of the
space-time dimension and of the covariant-gauge parameter $\xi$.
In ref.~\cite{DOT1}, they were obtained as a limiting case
of the general off-shell expressions. Of course, they may also
be obtained by direct calculation, using results for 
the on-shell one-loop integrals collected in Appendix~A of 
the present paper.
The corresponding one-loop expressions for the ghost-gluon
scalar functions are, for the on-shell limits of interest, collected
in Appendix~B. They can be obtained from the
expressions for general momenta presented in Appendix~D
of ref.~\cite{DOT1}, or by direct calculation.
The one-loop expressions for the two-point functions can be
found, for instance, in ref.~\cite{Muta} (see also in \cite{DOT1} and
Appendix~C of the present paper).
Collecting all the mentioned one-loop results,
we have checked that they satisfy
the conditions (\ref{WST1-1})--(\ref{WST-ghg}),   
in any space-time dimension $n$ and for any $\xi$.

In the one-loop expressions, the following
notation is used\footnote{Below, we shall also use the factor
$\eta$ in two-loop results (they are proportional to $\eta^2$). 
Since the two-loop contributions involve poles up to $1/\ep^4$, 
we need the expansion of $\eta$ up to the $\ep^4$ term.}:
\begin{equation}
\label{kappa}   
\kappa(p^2) \equiv
- \frac{2}{(n-3) (n-4)} \; (-p^2)^{(n-4)/2}
= \frac{1}{\varepsilon (1-2\varepsilon)} \; (-p^2)^{-\varepsilon} ,
\end{equation}
\bea   
\label{eta}
\eta &\equiv&
\frac{\Gamma^2(\frac{n}{2}-1)}{\Gamma(n-3)} \;
     \Gamma(3-{\textstyle{n\over2}}) =
\frac{\Gamma^2(1-\varepsilon)}{\Gamma(1-2\varepsilon)} \;
\Gamma(1+\varepsilon)
\nn \\
&=& e^{-\gamma\ep}\left[ 1-\frac{1}{12}\pi^2\ep^2 
-\frac{7}{3}\zeta_3\ep^3 
-\frac{47}{1440}\pi^4\ep^4
+ {\cal{O}}(\ep^5)
\right],
\eea
where $\gamma\simeq 0.57721566...$ is the Euler constant,
whilst $\zeta_3\equiv\zeta(3)=\sum_{j=1}^{\infty} j^{-3} \simeq
1.2020569...$
is the value of Riemann's zeta function.

We also use the standard notations $C_A$ for the eigenvalue
of the quadratic Casimir operator in the adjoint representation,
\begin{equation}
\label{C_A}
f^{acd}f^{bcd} = C_A \, \delta^{ab} \hspace{5mm}
(C_A = N \; \mbox{for the SU($N$) group}) ,
\end{equation}
and $C_F$ for the eigenvalue of the quadratic
Casimir operator in the fundamental representation. 
For the $\mbox{SU}(N)$ group, $C_F=(N^2-1)/(2N)$.
Furthermore,
\begin{equation}
\label{T_R}
T\equiv N_f T_R, \hspace{7mm}
T_R = {\textstyle{1\over8}} \; \mbox{Tr}(I) = {\textstyle{1\over2}} \; ,
\end{equation}
where $N_f$ is the number of quark flavours, and
$I$ is the ``unity'' in the space of Dirac matrices 
(we assume that $\mbox{Tr}(I)=4$).

\section{Planar two-loop three-point integrals}
\setcounter{equation}{0}

To calculate two-loop contributions to the three-gluon
vertex (shown in Fig.~1 of ref.~\cite{DOT2}), we consider
contractions of eq.~(\ref{ggg-U}) with all possible tensor
structures carrying three Lorentz indices $\mu_1, \mu_2$ and $\mu_3$. 
Note that non-planar graphs do not contribute to the two-loop
vertex, since their over-all colour factors vanish, due to the
Jacobi identity (see Fig.~6 of ref.~\cite{Cvit}, where this
is explained).

Technically, the problem is therefore reduced to the calculation 
of scalar integrals corresponding to the planar two-loop 
vertex graph shown in Fig.~3a (and similar graphs with
cyclic permutation of external momenta $p_1$, $p_2$ and $p_3$).
However, as a result of contracting the tensor structures, 
we get in the numerator some polynomials in scalar
products of external and loop momenta. The complete basis
for expanding these polynomials (see ref.~\cite{UD4}) 
includes (i) three external
momentum invariants (e.g.\ $p_1^2$, $p_2^2$ and $p_3^2$),
(ii) six squared momenta corresponding to the six denominators   
shown in Fig.~3a, and (iii) one additional invariant, which can
be chosen as $q^2$. 
Diagrammatically, the latter member of the basis can be associated
with the seventh line of an auxiliary ``forward-scattering'' 
four-point diagram shown in Fig.~3b. 

Since $q^2$ is missing in the original set of 
denominators (Fig.~3a), it always remains as a numerator,
which cannot be cancelled against any of the denominators involved.
Therefore, it is referred to as an {\em irreducible} numerator
(see, for instance, ref.~\cite{UD4}). 
In general, integrals with irreducible numerators require
a special consideration \cite{UD4,Tarasov}. However, as we shall see
below, this problem is not so serious in the on-shell case as
in the general off-shell case, since the relevant ``boundary''
integrals can be calculated for any (integer) powers of 
the numerator $q^2$.

In terms of an algorithm,
it is convenient to consider $q^2$ as an extra
denominator, remembering that its power $\nu_7$ is usually 
non-positive. Thus, let us consider integrals corresponding to
the auxiliary diagram in Fig.~3b:
\bea
\label{def_K}
K_3(n; \nu_1, \nu_2, \nu_3, \nu_4, \nu_5, \nu_6, \nu_7)
\hspace{96mm}
\nn \\
\equiv \int\int
\frac{\mbox{d}^n q \; \mbox{d}^n r}
     {[(p_1+r)^2]^{\nu_1}
      [(p_1+q)^2]^{\nu_2}
      [(p_2-r)^2]^{\nu_3}
      [(p_2-q)^2]^{\nu_4}
      (r^2)^{\nu_5}
      [(q-r)^2]^{\nu_6}
      (q^2)^{\nu_7} },
\eea
where $n=4-2\ep$ is the space-time dimension.
We shall also need diagrams corresponding to the permutations of the
external momenta in the diagram (\ref{def_K}). They are:
\bea
\label{def_K_2}
K_2(n; \nu_1, \nu_2, \nu_3, \nu_4, \nu_5, \nu_6, \nu_7)
\equiv \; \mbox{Eq.~(\ref{def_K})}\;\;
\mbox{with}\;\; (p_1, p_2, p_3) \rightarrow (p_3, p_1, p_2) , 
\\
\label{def_K_1}
K_1(n; \nu_1, \nu_2, \nu_3, \nu_4, \nu_5, \nu_6, \nu_7)
\equiv \; \mbox{Eq.~(\ref{def_K})}\;\;
\mbox{with}\;\; (p_1, p_2, p_3) \rightarrow (p_2, p_3, p_1) .
\eea
The corresponding diagrams are the same as given in Fig.~3;
the only thing to do is to permute the external momenta $p_i$. 
Note that for the integrals $K_2$ and $K_1$ not all external
lines of the four-point function are on shell, since some
of them carry $p_3$ with $p_3^2\equiv p^2 \neq 0$.
Instead, one of the corresponding Mandelstam variables vanishes.

To construct a procedure of calculating the integrals $K_i$
with different integer $\nu_i$, the integration-by-parts
procedure \cite{ibp} is useful\footnote{The application
of this procedure to the calculation of the integral 
$K_3(n;1,1,1,1,1,1,0)$ was presented in \cite{KL}
(see also eq.~(\ref{planar3}) below). In a more general context,
it was also discussed in \cite{StTk}.}.
If we introduce the notation ${\bf j}^+$ to denote an
{\it increase} of $\nu_j$ by one unit (and similarly let 
${\bf j}^-$ denote a {\it decrease} of $\nu_j$ by one unit), 
then the set of independent integration-by-parts relations 
for the functions (\ref{def_K}) can (for arbitrary momenta)
be written as follows\footnote{The set of recurrence
relations for calculating one-loop three-point functions
was considered in ref.~\cite{JPA}.}:
\be
\label{ibp1}
\left[ p_3^2 \nu_3 {\bf 3}^+ + \!p_1^2 \nu_5 {\bf 5}^+ 
       - \!\nu_3 {\bf 1}^- {\bf 3}^+ - \!\nu_5 {\bf 1}^- {\bf 5}^+
       + \!\nu_6 {\bf 6}^+ \left( {\bf 2}^- - {\bf 1}^- \right)
       + \!\left(n\!-\!2\nu_1\!-\!\nu_3\!-\!\nu_5\!-\!\nu_6\right)
\right] K_3 = 0, 
\ee
\be
\label{ibp2}
\left[ p_3^2 \nu_1 {\bf 1}^+ + \!p_2^2 \nu_5 {\bf 5}^+
       - \!\nu_1 {\bf 1}^+ {\bf 3}^- - \!\nu_5 {\bf 3}^- {\bf 5}^+
       + \!\nu_6 {\bf 6}^+ \left( {\bf 4}^- - {\bf 3}^- \right)
       + \!\left(n\!-\!\nu_1\!-\!2\nu_3\!-\!\nu_5\!-\!\nu_6\right)
\right] K_3 = 0,
\ee
\be
\label{ibp3}
\left[ p_1^2 \nu_1 {\bf 1}^+ + \!p_2^2 \nu_3 {\bf 3}^+
       - \!\nu_1 {\bf 1}^+ {\bf 5}^- - \!\nu_3 {\bf 3}^+ {\bf 5}^-
       + \!\nu_6 {\bf 6}^+ \left( {\bf 7}^- - {\bf 5}^- \right)
       + \!\left(n\!-\!\nu_1\!-\!\nu_3\!-\!2\nu_5\!-\!\nu_6\right)
\right] K_3 = 0,
\ee
\be
\label{ibp4}
\left[ \nu_1 {\bf 1}^+ \left( {\bf 2}^- - {\bf 6}^- \right)
     + \nu_3 {\bf 3}^+ \left( {\bf 4}^- - {\bf 6}^- \right)
     + \nu_5 {\bf 5}^+ \left( {\bf 7}^- - {\bf 6}^- \right)
     + \left(n\!-\!\nu_1\!-\!\nu_3\!-\!\nu_5\!-\!2\nu_6\right)
\right] K_3 = 0,
\ee
\be
\label{ibp5}
\left[ p_3^2 \nu_4 {\bf 4}^+ + \!p_1^2 \nu_7 {\bf 7}^+
       - \!\nu_4 {\bf 2}^- {\bf 4}^+ 
       + \!\nu_6 {\bf 6}^+ \left( {\bf 1}^- - {\bf 2}^- \right)
       - \!\nu_7 {\bf 2}^- {\bf 7}^+
       + \!\left(n\!-\!2\nu_2\!-\!\nu_4\!-\!\nu_6\!-\!\nu_7\right)
\right] K_3 = 0,
\ee
\be
\label{ibp6}
\left[ p_3^2 \nu_2 {\bf 2}^+ + \!p_2^2 \nu_7 {\bf 7}^+
       - \!\nu_2 {\bf 2}^+ {\bf 4}^- 
       + \!\nu_6 {\bf 6}^+ \left( {\bf 3}^- - {\bf 4}^- \right)
       - \!\nu_7 {\bf 4}^- {\bf 7}^+
       + \!\left(n\!-\!\nu_2\!-\!2\nu_4\!-\!\nu_6\!-\!\nu_7\right)
\right] K_3 = 0,
\ee
\be
\label{ibp7}
\left[ \nu_2 {\bf 2}^+ \left( {\bf 1}^- - {\bf 6}^- \right)
     + \nu_4 {\bf 4}^+ \left( {\bf 3}^- - {\bf 6}^- \right)
     + \nu_7 {\bf 7}^+ \left( {\bf 5}^- - {\bf 6}^- \right)
     + \left(n\!-\!\nu_2\!-\!\nu_4\!-\!2\nu_6\!-\!\nu_7\right)
\right] K_3 = 0,
\ee
\be
\label{ibp8}
\left[ p_1^2 \nu_2 {\bf 2}^+ + \!p_2^2 \nu_4 {\bf 4}^+
       - \!\nu_2 {\bf 2}^+ {\bf 7}^- - \!\nu_4 {\bf 4}^+ {\bf 7}^-
       + \!\nu_6 {\bf 6}^+ \left( {\bf 5}^- - {\bf 7}^- \right)
       + \!\left(n\!-\!\nu_2\!-\!\nu_4\!-\!\nu_6\!-\!2\nu_7\right)
\right] K_3 = 0.
\ee
Analogous relations for $K_2$ and $K_1$ 
can be obtained by permuting the subscripts of
$p_i^2$, according to eqs.~(\ref{def_K_2}) and (\ref{def_K_1}).

Now, if we recall that we are dealing with the on-shell case,
$p_1^2=p_2^2=0$, we can see that some terms on the r.h.s.\
of eqs.~(\ref{ibp1})--(\ref{ibp8}) (and in the analogous relations
for the  $K_2$ and $K_1$ integrals) vanish. 
In this case, the following symmetry property is 
valid:
\be
K_1(n;\nu_1,\nu_2,\nu_3,\nu_4,\nu_5,\nu_6,\nu_7)
=K_2(n;\nu_3,\nu_4,\nu_1,\nu_2,\nu_5,\nu_6,\nu_7)
\hspace{10mm} (\mbox{when}\; p_1^2=p_2^2).
\ee
Therefore, it is sufficient to consider the $K_3$ and $K_2$
integrals.

When certain powers of propagators $\nu_i$ 
vanish\footnote{Diagrammatically, this can be understood as shrinking
the corresponding line to a point.}, 
the corresponding integrals $K_i$ can be calculated 
in terms of $\Gamma$ functions or finite sums over terms
involving $\Gamma$ functions. 
A collection of relevant results for such ``boundary'' integrals
is presented in Appendix~D.
Using the relations (\ref{ibp1})--(\ref{ibp8}), the integrals $K_i$ 
with integer powers of propagators $\nu_i$ can be reduced to a set of
such ``boundary'' integrals, or to analogous integrals where some 
$\nu$'s are negative, i.e.\ the corresponding denominators are
in the numerator. 
The latter integrals can also be reduced to boundary integrals
(with the corresponding $\nu$'s equal to zero), 
by using the tensor decomposition in appropriate 
self-energy-type sub-loops, i.e.\ formulae similar to those
collected in Appendix~A of ref.~\cite{KL} (see also 
\cite{ibp,PLB'91}).

It should be noted that, starting from integrals with non-positive
$\nu_7$ and using relations (\ref{ibp1})--(\ref{ibp8}),
we never get integrals with positive $\nu_7$, since $\bf{7}^+$
is always accompanied by $\nu_7$ and one cannot ``overcome''
$\nu_7=0$. Nevertheless, in some cases integrals with 
positive $\nu_7$ may appear, due to the above-mentioned tensor 
decomposition in the sub-loops. Since the relevant boundary integrals
can be calculated for an arbitrary $\nu_7$ (see Appendix~D),
this does not create extra problems, even if all seven $\nu$'s
are positive (see eq.~(\ref{fsdb})).

Let us illustrate this procedure by some important examples.
One of them is the integral $K_3(1,1,1,1,1,1,0)$ (Fig.~3a), which 
was calculated in \cite{Gons,Neerven,KL}:
\bea
\label{planar3}
K_3(n;1,1,1,1,1,1,0)
\hspace{110mm}
\nn \\
= \frac{1}{2\ep^2}
\left[  K_3(n;0,2,1,0,1,2,0) - 2 K_3(n;0,1,0,2,2,1,0)
+ 2 K_3(n;1,1,2,2,0,0,0) \right]
\nn \\
= - \pi^{4-2\ep} (-p^2)^{-2-2\ep} \eta^2
\left[ \frac{1}{4\ep^4} 
+  \frac{1}{4\ep^2}\pi^2 
+ \frac{6}{\ep}\zeta_3 + \frac{3}{20}\pi^4
+ {\cal{O}}(\ep) \right].
\hspace{5mm}
\eea
To reduce the number of terms, some obvious symmetry properties 
of the boundary integrals have been used. 
Another example is the $K_2$ integral with the same $\nu$'s:
\bea
\label{planar2}
K_2(n;1,1,1,1,1,1,0)
\hspace{114mm}
\nn \\
=-\frac{1}{2\ep(1\!+\!2\ep)}
\left[ K_2(n;1,0,0,2,1,2,0)
\!+\! K_2(n;0,1,0,1,2,2,0) \!+\!2 K_2(n;1,0,0,3,1,1,0) \right]\!
\nn \\
= \pi^{4-2\ep} (-p^2)^{-2-2\ep} \eta^2
\left[ - \frac{3}{4\ep^3} + \frac{3}{2\ep^2} 
- \frac{3}{\ep} + \frac{1}{4\ep}\pi^2
+ 6 - \frac{1}{2}\pi^2 + 6\zeta_3 + {\cal{O}}(\ep) \right].
\hspace{5mm}
\eea
It corresponds to Fig.~3a, with the left lower momentum off shell. 
Expanding the $\eta$ factor (\ref{eta}), one can see that our results
(\ref{planar3})--(\ref{planar2}) coincide with those presented in 
\cite{Gons,Neerven,KL}.
Our eq.~(\ref{planar3}) corresponds to the result for diagram~6A
(with numerator $=1$) presented in the Appendix of ref.~\cite{Gons},
to the first line of Table~4 of ref.~\cite{Neerven} (``fig.~2''),
and to eq.~(12) of ref.~\cite{KL}.
Our eq.~(\ref{planar2}) corresponds to the second line of Table~4
of ref.~\cite{Neerven} (``fig.~4''),
and to eq.~(27) of ref.~\cite{KL}\footnote{In eq.~(27) of 
ref.~\cite{KL}, $9\zeta_4$ should read $9\zeta_2$. We also note
another obvious misprint: in the last diagram on the r.h.s.
of eq.~(9) of \cite{KL}, the ``upper'' line should contain a dot,
i.e.\ the power of the propagator is equal to 2, which is clear
from their eqs.~(8) and (10).}.  
We have also checked the results for other scalar integrals
listed in refs.~\cite{Gons,Neerven,KL}, namely: all results for
the diagrams from 6A to 3A in the Appendix of ref.~\cite{Gons}
(including those with numerators\footnote{We note a misprint
in the result for the diagram~4C with numerator $(l\cdot p)$,
$15/(16\omega)$ should read $5/(16\omega)$ (in our notation,
$\omega \leftrightarrow -\ep$). In addition, $l^2$ is
forgotten in the product of denominators in diagram~5F, 
and the denominator $(l-r)^2$ should read $(l+r)^2$ in
diagram~4D.}, but excluding the 
non-planar diagram~6B); the three remaining lines in Table~4
of ref.~\cite{Neerven}, as well as the results listed in
Appendix~C of \cite{Neerven}\footnote{In eq.~(C.5) of 
ref.~\cite{Neerven}, in the second term in the square brackets
(the term containing $D$) the numerator $2+\ep^2$ should
read $2+\textstyle{1\over2}\ep^2$ (note that the $\ep$ used in
\cite{Neerven}
corresponds to our $-2\ep$).} (excluding the non-planar
stuff, eqs.~(C.6) and (C.7)); eqs.~(26) and (28) of ref.~\cite{KL}.

The algorithm discussed in this section makes it possible to 
evaluate all relevant integrals (corresponding to planar two-loop
diagrams) for an {\em arbitrary} value of the space-time dimension
$n$. However, since in physical applications one usually needs
the expansion in $\ep=(4-n)/2$, we present below the results for the
two-loop vertices in an expanded form, up to $\ep^0$, i.e.\ keeping 
the poles and the finite terms.

\section{Two-loop results for the three-gluon vertex}
\setcounter{equation}{0}

Below, we list the unrenormalized two-loop contributions to the
functions $U_i(p^2)$ occurring in the three-gluon vertex.
They were calculated using a set of {\sf REDUCE} \cite{reduce} programs
based on the algorithm described in the previous
section. 
The two-loop diagrams contributing to the three-gluon vertex are
shown in Fig.~1 of ref.~\cite{DOT2}.
The results are expanded in $\ep$ up to the finite terms. 
The factor $\eta$ is defined (and its expansion in $\ep$ is given) 
in eq.~(\ref{eta}). The colour factors $C_A$, $T$ and $C_F$ are
defined at the end of section~2.

The contributions of the diagrams without quark loops, in an
arbitrary covariant gauge (for the definition of
the gauge parameter $\xi$, see eq.~(\ref{gl_prop})) are
\bea
U_1^{(2,\xi)}(p^2) & \!=\! & C_A^2 \; \frac{g^4\;\eta^2}{(4\pi)^n} 
(-p^2)^{-2\ep}
\left\{
\frac{1}{\ep^4}\left( \frac{21}{64} + \frac{9}{128}\xi \right)
+ \frac{1}{\ep^3}\left( \frac{11}{12}  + \frac{335}{384}\xi
                        +\frac{25}{256}\xi^2 \right)
\right.
\nn \\
&&+ \frac{1}{\ep^2}\left(- \frac{37}{72} + \frac{19}{192}\pi^2
      + \frac{623}{576}\xi + \frac{3}{128}\pi^2\xi
+ \frac{103}{128}\xi^2 + \frac{1}{32}\xi^3  \right)
\nn \\
&&+ \frac{1}{\ep}\left( 
- \frac{239}{54} + \frac{1}{72}\pi^2 + \frac{11}{4}\zeta_3
+ \frac{3343}{1728}\xi 
+ \frac{37}{384}\pi^2\xi
\right.
\nn \\
&& \left.
+ \frac{21}{32}\zeta_3\xi  
+ \frac{137}{384}\xi^2 - \frac{1}{256}\pi^2\xi^2
     + \frac{9}{16}\xi^3 \right)
\nn \\
&& - \frac{6653}{324} - \frac{119}{432}\pi^2 + \frac{139}{12}\zeta_3
+ \frac{21}{320}\pi^4
- \frac{1333}{5184}\xi + \frac{37}{192}\pi^2\xi + \frac{191}{32}\zeta_3\xi
\nn \\
&& \left.
+ \frac{1}{64}\pi^4\xi
+ \frac{1993}{1152}\xi^2 - \frac{3}{128}\pi^2\xi^2
- \frac{11}{64}\zeta_3\xi^2 
+ \frac{19}{16}\xi^3 + \frac{1}{16}\xi^4 
\right\} + {\cal{O}}(\ep),
\eea
\bea
U_2^{(2,\xi)}(p^2) & \!=\! & C_A^2 \; \frac{g^4\;\eta^2}{(4\pi)^n} 
(-p^2)^{-2\ep}
\left\{
- \frac{5}{16\ep^4}
- \frac{1}{\ep^3}\left( \frac{19}{16} +\frac{5}{8}\xi \right)
\right.
\nn \\
&& + \frac{1}{\ep^2}\left( \frac{15}{8} - \frac{5}{48}\pi^2 
- \frac{1}{12}\xi - \frac{1}{192}\pi^2\xi - \frac{61}{64}\xi^2
\right)
\nn \\
&& + \frac{1}{\ep}\left( \frac{593}{48} - \frac{1}{36}\pi^2 
- \frac{7}{4}\zeta_3 
- \frac{67}{288}\xi - \frac{1}{12}\pi^2\xi
- \frac{1}{32}\zeta_3\xi  + \frac{31}{48}\xi^2 
- \frac{13}{16}\xi^3 
\right)
\nn \\
&& + \frac{17939}{288} + \frac{19}{108}\pi^2 - \frac{107}{12}\zeta_3 
- \frac{1}{20}\pi^4 
+ \frac{8381}{1728}\xi - \frac{31}{96}\pi^2\xi - 5\zeta_3\xi 
\nn \\
&& \left.
- \frac{1}{640}\pi^4\xi
- \frac{41}{288}\xi^2 - \frac{1}{192}\pi^2\xi^2 
+ \frac{1}{16}\zeta_3\xi^2 - \frac{11}{8}\xi^3 - \frac{1}{8}\xi^4   
\right\} + {\cal{O}}(\ep),
\eea
\bea
U_3^{(2,\xi)}(p^2) & \!=\! & C_A^2 \; \frac{g^4\;\eta^2}{(4\pi)^n} 
(-p^2)^{-2\ep}
\left\{
\frac{1}{\ep^4}\left(  
\frac{1}{16} - \frac{1}{8}\xi - \frac{1}{32}\xi^2 \right) 
\right.
\nn \\
&& +\frac{1}{\ep^3}\left( \frac{79}{96} - \frac{65}{192}\xi 
   - \frac{103}{192}\xi^2  - \frac{27}{256}\xi^3 \right) 
\nn \\
&& + \frac{1}{\ep^2}\left( \frac{95}{72} + \frac{17}{48}\pi^2 
      + \frac{157}{288}\xi - \frac{13}{128}\pi^2\xi 
      - \frac{751}{1152}\xi^2 - \frac{1}{192}\pi^2\xi^2 
      - \frac{81}{256}\xi^3 - \frac{1}{32}\xi^4 \right)
\nn \\
&& + \frac{1}{\ep}\left( - \frac{6125}{864} + \frac{155}{72}\pi^2 
      + \frac{13}{4}\zeta_3
      + \frac{97}{1728}\xi - \frac{1}{48}\pi^2\xi  
      - \frac{135}{64}\zeta_3\xi
\right.
\nn \\
&&
\left.
      - \frac{1433}{864}\xi^2 + \frac{3}{128}\pi^2\xi^2 
      - \frac{1}{32}\zeta_3\xi^2 
      - \frac{29}{32}\xi^3 - \frac{1}{768}\pi^2\xi^3  
      - \frac{1}{8}\xi^4 \right)
\nn \\
&& - \frac{265205}{5184} + \frac{206}{27}\pi^2 
+ \frac{227}{12}\zeta_3 + \frac{1}{8}\pi^4 
+ \frac{140815}{10368}\xi - \frac{319}{288}\pi^2\xi 
- \frac{239}{32}\zeta_3\xi 
\nn \\
&& - \frac{71}{1280}\pi^4\xi 
- \frac{23687}{5184}\xi^2  + \frac{1}{6}\pi^2\xi^2 
- \frac{9}{64}\zeta_3\xi^2 - \frac{1}{640}\pi^4\xi^2
- \frac{145}{64}\xi^3  
\nn \\
&& \left.
- \frac{5}{768}\pi^2\xi^3 
+ \frac{5}{32}\zeta_3\xi^3 - \frac{13}{32}\xi^4 
\right\} 
+ {\cal{O}}(\ep),
\eea
\bea
p^2 U_4^{(2,\xi)}(p^2) & \!=\! & -C_A^2 \; \frac{g^4\;\eta^2}{(4\pi)^n} 
(-p^2)^{-2\ep}
\left\{
\frac{1}{\ep^4}\left( \frac{11}{32} + \xi 
       + \frac{55}{128}\xi^2 + \frac{5}{128}\xi^3 \right)
\right.
\nn \\ 
&& +\frac{1}{\ep^3}\left( \frac{11}{16} + \frac{301}{96}\xi 
       + \frac{539}{192}\xi^2 + \frac{65}{48}\xi^3 
       + \frac{25}{128}\xi^4 \right)
\nn \\
&& +\frac{1}{\ep^2}\left( \frac{9}{16} + \frac{11}{96}\pi^2 
- \frac{389}{144}\xi + \frac{9}{16}\pi^2\xi
+ \frac{1541}{1152}\xi^2 + \frac{49}{192}\pi^2\xi^2 
\right.
\nn \\
&& \left.
+ \frac{4267}{2304}\xi^3 
+ \frac{1}{32}\pi^2\xi^3 + \frac{3}{16}\xi^4  
+ \frac{1}{16}\xi^5 \right)
\nn \\
&& +\frac{1}{\ep}\left( \frac{2195}{96} \!-\! \frac{271}{144}\pi^2 
\!+\! \frac{19}{8}\zeta_3 
\!-\! \frac{34549}{1728}\xi \!+\! \frac{7}{9}\pi^2\xi 
\!+\! \frac{15}{2}\zeta_3\xi
\!+\! \frac{7031}{1728}\xi^2 \!-\! \frac{5}{576}\pi^2\xi^2 
\right.
\nn \\
&& \left.
+ \frac{221}{64}\zeta_3\xi^2
+ \frac{20881}{3456}\xi^3 - \frac{1}{16}\pi^2\xi^3  
+ \frac{21}{64}\zeta_3 \xi^3
+ \frac{293}{256}\xi^4 + \frac{1}{384}\pi^2\xi^4 
+ \frac{3}{32}\xi^5
\right)
\nn \\
&& + \frac{84683}{576} - \frac{2473}{216} \pi^2 
+ \frac{89}{24} \zeta_3 + \frac{1}{16} \pi^4
- \frac{1212661}{10368} \xi + \frac{685}{216}\pi^2 \xi
\nn \\
&& + \frac{1661}{48} \zeta_3 \xi
+ \frac{19}{80} \pi^4 \xi
- \frac{39857}{5184} \xi^2 + \frac{1327}{3456} \pi^2 \xi^2
+ \frac{109}{6} \zeta_3 \xi^2 
\nn \\
&& + \frac{139}{1280} \pi^4 \xi^2
+ \frac{87085}{5184} \xi^3 + \frac{73}{768} \pi^2 \xi^3
- \frac{21}{32} \zeta_3 \xi^3
+ \frac{3}{256} \pi^4 \xi^3 
\nn \\
&& \left.
+ \frac{89}{32} \xi^4 + \frac{1}{192} \pi^2 \xi^4 
- \frac{7}{32} \zeta_3 \xi^4  
+ \frac{7}{32} \xi^5
\right\} + {\cal{O}}(\ep),
\eea
\bea  
p^2 U_5^{(2,\xi)}(p^2) & \!=\! & C_A^2 \; \frac{g^4\;\eta^2}{(4\pi)^n} 
(-p^2)^{-2\ep}
\left\{
-\frac{1}{\ep^4}\left(  \frac{1}{8} + \frac{9}{64} \xi \right)
+ \frac{1}{\ep^3}\left( \frac{1}{24} - \frac{155}{192} \xi 
          - \frac{25}{128} \xi^2  \right)
\right.
\nn \\
&& + \frac{1}{\ep^2}\left(  \frac{10}{9}  - \frac{1}{16} \pi^2
- \frac{653}{288} \xi - \frac{7}{192} \pi^2\xi 
 - \frac{67}{128} \xi^2 - \frac{1}{16} \xi^3 \right)
\nn \\
&& + \frac{1}{\ep}\left( \frac{385}{216} \!-\! \frac{11}{72} \pi^2
\!-\! \frac{21}{8} \zeta_3 \!-\! \frac{4303}{864} \xi 
\!-\! \frac{5}{192} \pi^2\xi 
\!-\! \frac{5}{4} \zeta_3 \xi
\!-\! \frac{63}{64} \xi^2  \!+\! \frac{1}{128} \pi^2\xi^2
\!\!-\! \frac{5}{16} \xi^3\!
\right)
\nn \\
&& - \frac{9557}{1296} - \frac{41}{108}\pi^2\!
- \frac{113}{12} \zeta_3 - \frac{9}{160} \pi^4\! 
- \frac{12889}{1296} \xi + \frac{9}{32}\pi^2\xi
- \frac{103}{16} \zeta_3 \xi - \frac{9}{320} \pi^4 \xi
\nn \\
&& \left. 
- \frac{457}{192} \xi^2 + \frac{7}{128} \pi^2 \xi^2
+ \frac{9}{32} \zeta_3 \xi^2 - \xi^3
\right\} + {\cal{O}}(\ep),
\eea
\bea  
p^2 U_6^{(2,\xi)}(p^2) & \!=\! & -C_A^2 \; \frac{g^4\;\eta^2}{(4\pi)^n} 
(-p^2)^{-2\ep}
\left\{
\frac{3}{32\ep^4} \xi 
+ \frac{1}{\ep^3}\left( - \frac{1}{3} + \frac{41}{192} \xi 
+ \frac{41}{128} \xi^2 \right) 
\right.
\nn \\
&& + \frac{1}{\ep^2}\left( \frac{157}{144} - \frac{1}{48} \pi^2
- \frac{53}{144} \xi  + \frac{9}{64} \pi^2\xi 
 - \frac{155}{128} \xi^2 + \frac{13}{384} \pi^2 \xi^2
+ \frac{29}{256} \xi^3 \right)
\nn \\
&& + \frac{1}{\ep}\left( \frac{2365}{864}- \frac{1}{36} \pi^2
- \frac{1}{8} \zeta_3
- \frac{3565}{1728} \xi + \frac{5}{64}  \pi^2 \xi
+ \frac{33}{32} \zeta_3 \xi
\right.
\nn \\
&& \left.
- \frac{73}{192} \xi^2 + \frac{11}{128} \pi^2\xi^2
+ \frac{13}{64} \zeta_3 \xi^2
- \frac{27}{64} \xi^3 \right)
\nn \\
&& + \frac{40429}{5184}- \frac{47}{432} \pi^2\!
+ \frac{31}{12} \zeta_3 - \frac{1}{160} \pi^4\!
- \frac{150367}{10368} \xi + \frac{269}{288} \pi^2\xi
- \frac{3}{2} \zeta_3 \xi + \frac{29}{640}  \pi^4\xi
\nn \\
&& \left.
- \frac{29}{9} \xi^2 \!+\! \frac{113}{384} \pi^2\xi^2
\!+\! \frac{1}{2} \zeta_3 \xi^2 \!+\! \frac{13}{1280} \pi^4\xi^2
\!-\! \frac{3}{4} \xi^3 \!-\! \frac{1}{768} \pi^2\xi^3 
\!-\! \frac{3}{32} \xi^4\!
\right\} \!+\! {\cal{O}}(\ep),
\hspace{3mm}
\eea
\bea
p^2 U_7^{(2,\xi)}(p^2) & \!=\! & C_A^2 \; \frac{g^4\;\eta^2}{(4\pi)^n} 
(-p^2)^{-2\ep}
\left\{
\frac{1}{\ep^4}\left( \frac{21}{32} - \frac{5}{32} \xi - \frac{1}{16} \xi^2
\right)
\right.
\nn \\
&& +\frac{1}{\ep^3}\left(  \frac{193}{48}+ \frac{139}{192} \xi
- \frac{289}{384} \xi^2 - \frac{27}{128} \xi^3 \right)
\nn \\
&&
+\frac{1}{\ep^2}\left( \frac{125}{144}+ \frac{83}{96} \pi^2\!
+ \frac{119}{144} \xi - \frac{1}{24} \pi^2 \xi
- \frac{355}{576} \xi^2\! + \frac{7}{384} \pi^2\xi^2
- \frac{133}{256} \xi^3\! - \frac{1}{16} \xi^4 \right)
\nn \\ 
&& +\frac{1}{\ep}\left( - \frac{29137}{864} + \frac{311}{72} \pi^2
+ \frac{19}{2} \zeta_3
- \frac{2027}{1728} \xi + \frac{43}{192} \pi^2\xi
- \frac{49}{16} \zeta_3 \xi
\right.
\nn \\
&& \left.
- \frac{8621}{1728} \xi^2 + \frac{43}{384} \pi^2\xi^2
+\frac{7}{64} \zeta_3 \xi^2 
- \frac{39}{64} \xi^3- \frac{1}{384} \pi^2\xi^3 - \frac{1}{4} \xi^4 \right)
\nn \\
&& - \frac{1100617}{5184} + \frac{6377}{432} \pi^2 
+ \frac{337}{6} \zeta_3 + \frac{53}{160} \pi^4
+ \frac{50779}{10368} \xi - \frac{19}{32} \pi^2 \xi
\nn \\
&& - \frac{107}{16} \zeta_3 \xi - \frac{19}{320}  \pi^4\xi
- \frac{15529}{1296} \xi^2 + \frac{55}{96} \pi^2 \xi^2
- \frac{1}{32} \zeta_3 \xi^2 + \frac{7}{1280}  \pi^4 \xi^2 
\nn \\ 
&& \left. 
- \frac{81}{32} \xi^3
- \frac{11}{768} \pi^2 \xi^3
+ \frac{5}{16} \zeta_3 \xi^3 
 - \frac{21}{32} \xi^4 
\right\} 
+ {\cal{O}}(\ep).
\eea

The unrenormalized two-loop contributions of the diagrams involving
quark loops are
\bea
U_1^{(2,q)}(p^2) & \!=\! & C_A T \; \frac{g^4\;\eta^2}{(4\pi)^n}
(-p^2)^{-2\ep}
\left\{
-\frac{1}{\ep^3} \left( \frac{17}{24} + \frac{1}{6} \xi \right)
+ \frac{1}{\ep^2} \left( \frac{17}{18} - \frac{29}{18} \xi \right)
\right.
\nn \\
&& \left.
+ \frac{1}{\ep} \left(  \frac{383}{108} + \frac{1}{72} \pi^2
 - \frac{85}{54} \xi - \frac{2}{3} \xi^2 \right)
- \frac{449}{324} + \frac{43}{108} \pi^2 + \frac{19}{3} \zeta_3
- \frac{197}{162} \xi - \frac{41}{18} \xi^2
\right\}
\nn \\
&& +C_F T\;  \frac{g^4\;\eta^2}{(4\pi)^n}
(-p^2)^{-2\ep}
\left\{
\frac{2}{\ep} + \frac{61}{3} - 16 \zeta_3 \right\}
+ {\cal{O}}(\ep),
\eea
\bea
U_2^{(2,q)}(p^2) & \!=\! & C_A T \; \frac{g^4\;\eta^2}{(4\pi)^n}
(-p^2)^{-2\ep}
\left\{
\frac{1}{\ep^3} 
+ \frac{1}{\ep^2}\left(- \frac{3}{2}+\frac{5}{3}\xi \right)
\right.
\nn \\
&& \left.
- \frac{1}{\ep}\left( \frac{91}{12} + \frac{1}{36} \pi^2
+ \frac{5}{9} \xi - \frac{4}{3} \xi^2 \right)
- \frac{2113}{72} - \frac{23}{108} \pi^2 - \frac{85}{6} \zeta_3
- \frac{190}{27} \xi + \frac{44}{9} \xi^2 
\right\}
\nn \\
&& +C_F T\;  \frac{g^4\;\eta^2}{(4\pi)^n}
(-p^2)^{-2\ep}
\left\{
- \frac{4}{\ep} - \frac{110}{3} + 32 \zeta_3 \right\}
+ {\cal{O}}(\ep),
\eea
\bea
U_3^{(2,q)}(p^2) & \!=\! & C_A T \; \frac{g^4\;\eta^2}{(4\pi)^n}
(-p^2)^{-2\ep}
\left\{
\frac{1}{\ep^3}
\left( - \frac{7}{12}  + \frac{1}{3} \xi + \frac{1}{6} \xi^2 \right)
\right.
\nn \\
&& + \frac{1}{\ep^2}
\left( - \frac{35}{36} - \frac{1}{6} \pi^2 + \frac{1}{18} \xi 
       + \frac{11}{18} \xi^2 \right)
+ \frac{1}{\ep}
\left( \frac{4405}{216} - \frac{151}{72} \pi^2 \!- \zeta_3 
\!- \frac{23}{108} \xi \!+ \frac{47}{27} \xi^2 \right)
\nn \\
&& \left.
+ \frac{152755}{1296}  - \frac{1661}{216} \pi^2 - \frac{133}{12} \zeta_3
- \frac{1}{20} \pi^4 - \frac{235}{324} \xi + \frac{1}{36} \pi^2 \xi
+ 2 \zeta_3 \xi + \frac{337}{81} \xi^2
\right\}
\nn \\
&& +C_F T\;  \frac{g^4\;\eta^2}{(4\pi)^n}
(-p^2)^{-2\ep}
\left\{ 
- \frac{1}{3\ep} \pi^2 
+ \frac{26}{3} - 2 \pi^2 - 10 \zeta_3 \right\}
+ {\cal{O}}(\ep),
\eea
\bea   
p^2 U_4^{(2,q)}(p^2) & \!=\! & C_A T \; \frac{g^4\;\eta^2}{(4\pi)^n}
(-p^2)^{-2\ep}   
\left\{
\frac{1}{\ep^3}
\left(
\frac{5}{4} + \frac{29}{24} \xi + \frac{19}{12} \xi^2 + \frac{1}{3} \xi^3
\right)
\right.
\nn \\
&& + \frac{1}{\ep^2}
\left( \frac{103}{4} - \frac{1}{3} \pi^2- \frac{203}{72} \xi
+ \frac{73}{18} \xi^2 + \frac{7}{18} \xi^3 \right)
\nn \\
&& + \frac{1}{\ep}
\left(  \frac{5125}{24} - \frac{53}{9} \pi^2 - 2 \zeta_3 
- \frac{457}{54} \xi - \frac{17}{72} \pi^2 \xi
 + \frac{1111}{108} \xi^2  - \frac{1}{36} \pi^2 \xi^2 
+ \frac{25}{27} \xi^3 \right)
\nn \\
&& + \frac{153277}{144} - \frac{3307}{108} \pi^2 
- \frac{113}{6} \zeta_3
- \frac{1}{10} \pi^4 - \frac{3346}{81} \xi - \frac{31}{216} \pi^2 \xi
 + \frac{28}{3} \zeta_3 \xi  
\nn \\
&& \left.
+ \frac{1919}{81} \xi^2
+ \frac{1}{27} \pi^2 \xi^2 + \frac{7}{3} \zeta_3 \xi^2
+ \frac{131}{81} \xi^3 \right\}
\nn \\
&& - C_F T\;  \frac{g^4\;\eta^2}{(4\pi)^n}
(-p^2)^{-2\ep}
\left\{
\frac{8}{\ep^2} \!+\! \frac{1}{\ep}\left( 52 \!+\! \frac{2}{3} \pi^2\! \right)
\!+\! 214 \!+\! \frac{4}{3} \pi^2 \!+\! 52 \zeta_3\! \right\}
\!+\! {\cal{O}}(\ep), \hspace{8mm}
\eea
\bea
p^2 U_5^{(2,q)}(p^2) & \!=\! & C_A T \; \frac{g^4\;\eta^2}{(4\pi)^n}
(-p^2)^{-2\ep}
\left\{ 
\frac{1}{\ep^3}
\left( - \frac{1}{12} + \frac{1}{3} \xi \right)
+ \frac{1}{\ep^2}
\left( - \frac{14}{9} + \frac{14}{9} \xi \right)
\right.
\nn \\
&& \left.
+ \frac{1}{\ep}
\left( - \frac{64}{27} - \frac{1}{36} \pi^2 + \frac{100}{27} \xi \right)
 + \frac{8101}{324} - \frac{37}{54} \pi^2 + \frac{1}{3} \zeta_3
+ \frac{767}{81} \xi - \frac{1}{3} \xi^2 
\right\}
\nn \\
&& -4 C_F T\;  \frac{g^4\;\eta^2}{(4\pi)^n} (-p^2)^{-2\ep}
+ {\cal{O}}(\ep), \hspace{8mm}
\eea
\bea
p^2 U_6^{(2,q)}(p^2) & \!=\! & C_A T \; \frac{g^4\;\eta^2}{(4\pi)^n}
(-p^2)^{-2\ep}
\left\{
\frac{1}{\ep^3}
\left( - \frac{2}{3} + \frac{11}{24} \xi \right)
+ \frac{1}{\ep^2}
\left(  \frac{59}{36} - \frac{65}{72} \xi \right)
\right.
\nn \\
&& + \frac{1}{\ep}
\left(  \frac{2387}{216} - \frac{13}{18} \pi^2 + \frac{49}{27} \xi
- \frac{1}{24}  \pi^2 \xi  - \frac{7}{6} \xi^2 \right)
\nn \\
&& \left.
+ \frac{42515}{1296} - \frac{223}{108} \pi^2 + \frac{35}{3} \zeta_3 
+ \frac{1519}{162} \xi - \frac{11}{72} \pi^2\xi 
+ \zeta_3 \xi - \frac{77}{18} \xi^2 
\right\}
\nn \\
&& + C_F T\;  \frac{g^4\;\eta^2}{(4\pi)^n}
(-p^2)^{-2\ep}
\left\{
\frac{4}{\ep} + \frac{110}{3} - 32 \zeta_3
\right\}
+ {\cal{O}}(\ep),
\eea
\bea
p^2 U_7^{(2,q)}(p^2) & \!=\! & C_A T \; \frac{g^4\;\eta^2}{(4\pi)^n}
(-p^2)^{-2\ep}
\left\{ 
\frac{1}{\ep^3}  
\left(- \frac{8}{3} + \frac{5}{24} \xi + \frac{1}{3} \xi^2 \right) 
\right.
\nn \\
&& +\frac{1}{\ep^2}
\left( - \frac{37}{36} - \frac{1}{3} \pi^2 - \frac{167}{72} \xi
+ \frac{11}{9} \xi^2 \right)
\nn \\ 
&& +\frac{1}{\ep}
\left( 
\frac{9419}{216} - \frac{125}{36} \pi^2  - 2 \zeta_3 
- \frac{61}{54} \xi  + \frac{1}{24}  \pi^2  \xi
+ \frac{107}{54} \xi^2 \right)
\nn \\
&& \left.
+ \frac{334055}{1296} - \frac{352}{27} \pi^2 - \frac{29}{6} \zeta_3
- \frac{1}{10} \pi^4 + \frac{263}{81} \xi + \frac{5}{24} \pi^2  \xi
+ 3 \zeta_3 \xi + \frac{457}{162} \xi^2
\right\}
\nn \\
&& + C_F T\;  \frac{g^4\;\eta^2}{(4\pi)^n}
(-p^2)^{-2\ep}
\left\{
\frac{1}{\ep} \left( 4 - \frac{2}{3} \pi^2 \right)
+ 54 - 4 \pi^2 - 52 \zeta_3 \right\}
+ {\cal{O}}(\ep).
\eea

We have also obtained the two-loop results
for the ghost-gluon scalar functions
in all on-shell limits of interest. They are given in Appendix~E,
whereas the relevant two-loop contributions to the two-point
functions $J(p^2)$ and $G(p^2)$ are collected in Appendix~C.
Using all these expressions, together with the one-loop contributions,
we have checked that all the results obtained satisfy the
WST identities (\ref{WST1-1})--(\ref{WST-ghg}), as they should.

The renormalization of the results for the three-gluon vertex
(and other three- and two-point functions involved) was discussed 
in detail in section~8 of ref.~\cite{DOT2}. 
The corresponding renormalization factors ($Z_1$, $\widetilde{Z_1}$,
$Z_3$ and  $\widetilde{Z_3}$) in the MS (or $\overline{\mbox{MS}}$) scheme
have been presented in refs.~\cite{Z-factors,PT-book}. 
For a detailed discussion of these results, 
together with a list of misprints,
see Appendix~B of ref.~\cite{DOT2}. 
To construct renormalized expressions for the three-gluon vertex
functions $U_i$ at the two-loop level, we need \\
(i) to take the sum of (unrenormalized) zero-, one- and two-loop
contributions\footnote{Note that the one-loop expressions should be
expanded up to $\ep^2$ terms, since they may be multiplied by other
one-loop contributions involving $1/\ep^2$ poles.}, considering
the coupling constant and the gauge 
parameter as ``bare'' quantities, $g\rightarrow g_{\rm B}$ and 
$\xi\rightarrow \xi_{\rm B}$; \\
(ii) to substitute $g_{\rm B}$ and $\xi_{\rm B}$ in terms of the renormalized
$g$ and $\xi$, multiplied by the appropriate $Z$-factors 
(see eqs.~(8.8) and (8.9) of ref.~\cite{DOT2}); \\
(iii) to multiply the resulting expression by the corresponding
$Z$-factor\footnote{For the ghost-gluon vertex (see Appendix~E),
the renormalization factor $\widetilde{Z}_1$ is required
(see, for instance, eq.~(B.2) in ref.~\cite{DOT2}).}, 
namely $Z_1$ (see eq.~(B.1) of ref.~\cite{DOT2}). 

Since the resulting renormalized expressions are as cumbersome
as the unrenormalized ones (and can easily be obtained from the
latter ones), we do not present them here. They also contain
infrared (on-shell) poles in $\ep$ up to $1/\ep^4$.

\section{Conclusion}
\setcounter{equation}{0}

In the limit when two external gluons are on shell,
we have calculated the two-loop contributions to the three-gluon
vertex, in an arbitrary covariant gauge, keeping
finite terms of the expansion in $\ep=(4-n)/2$.
In this limit, the three-gluon vertex is described by
seven scalar functions $U_i(p^2)$
associated with different tensor structures, see eq.~(\ref{ggg-U}).
The results (listed in section~4) contain on-shell singularities
up to $1/\ep^4$. The ultraviolet singularities are at most $1/\ep^2$
and should be removed by the renormalization. In a realistic
physical calculation of squared amplitudes, the infrared
(on-shell) singularities
should be cancelled by the contributions of one-loop diagrams with 
soft emission from the external legs, etc., according to the
Kinoshita--Lee--Nauenberg mechanism \cite{KNL}.
In this way, our result will be useful as a ``block'' in the calculation 
of NNLO corrections to physical amplitudes.

We have also calculated the ghost-gluon vertex (\ref{BC-ghg})
in all on-shell limits of interest; the results are
collected in Appendix~E.
We have confirmed that the obtained results obey the 
corresponding WST identities (\ref{WST1-1})--(\ref{WST-ghg}).

We note that in the on-shell case considered, the problem
of irreducible numerators in three-point two-loop integrals already 
shows up, but it can be overcome in a relatively simple way,
since the relevant boundary integrals can be calculated
for any integer powers of this numerator (see Appendix~D).
In the zero-momentum calculation \cite{DOT2}, there was no such problem
at all. In the general off-shell calculation, though, the problem
of irreducible numerators is much more severe \cite{UD4}.

Our results can be considered as a further step,
in addition to \cite{BL,DOT2}, towards calculating the two-loop
QCD vertices in more complicated cases,
such as the on-shell limit with just one gluon on shell,
or the general off-shell case\footnote{In principle, the techniques
for calculating the relevant off-shell scalar integrals
are already available \cite{UD13,UD4}.}.   

\vspace{10mm}  

{\bf Acknowledgements.}
We are grateful to S.~Catani, W.L. van Neerven and O.V.~Tara\-sov
for useful discussions.
A.~D. is grateful to the Department of Physics, University of Bergen
(where this work was started), and also to 
the Department of Physics, University of Mainz (where this work 
was completed), for their hospitality.
P.~O. would like to thank the Theory groups of DESY and CERN
for kind hospitality.
This research has been essentially supported by the Research Council 
of Norway and the Alexander von Humboldt Foundation (A.~D.).   
A partial support from the grants RFBR-98-02-16981 and 
Volkswagen--I/73611 is acknowledged (A.~D.).

\newpage

\section*{Appendix A: Some useful one-loop formulae}
\setcounter{equation}{0}
\renewcommand{\thesection}{A}

The ``triangle'' integral with the external momenta $p_1$, $p_2$ and
$p_3=-p_1-p_2$ is defined as
\be 
\label{def_J}
J(n;\nu_1,\nu_2,\nu_3)\equiv \int
\frac{\mbox{d}^n q}
     {[(p_2-q)^2]^{\nu_1} [(p_1+q)^2]^{\nu_2} (q^2)^{\nu_3}} ,
\ee
where $n=4-2\ep$ is the space-time dimension.
For general values of $n$ and $\nu_i$, the result for the
integral (\ref{def_J}) can be expressed in terms of 
hypergeometric functions of two variables \cite{BD-VMU,JPA}. 

When two external legs are on shell, $p_1^2=p_2^2=0$ ($p_3^2\equiv p^2$),
the following simple formula can be easily obtained:
\be
\label{os_triang}
\left. \frac{}{} \!\! J(n;\nu_1,\nu_2,\nu_3)
\right|_{\begin{array}{c} {}_{\!\!p_1^2=p_2^2=0} \\
                          {}^{\!\!p_3^2\equiv p^2}
         \end{array}}
= \mbox{i}^{1-n} \pi^{n/2} (p^2)^{n/2-\Sigma\nu_i} \;
\frac{\Gamma\left(\halfn\!-\!\nu_1\!-\!\nu_3\right)
      \Gamma\left(\halfn\!-\!\nu_2\!-\!\nu_3\right)
      \Gamma\left(\sum\nu_i\!-\!\halfn\right)}
     {\Gamma(\nu_1) \Gamma(\nu_2) \Gamma\left(n-\sum\nu_i\right)} .
\ee
In particular, we shall need this formula for the case when
one of the indices is a negative integer, $\nu_3=-s$.
This means that $(q^2)^s$ is in the numerator.
When $\nu_3=0$, the r.h.s. of (\ref{os_triang}) gives the
well-known result for the one-loop two-point function. 
When $\nu_1$ or $\nu_2$ are non-positive integers, we get zero.

We also need the result for the triangle integral with one leg 
on shell. Assuming that $p_1^2\neq 0, \; p_2^2\neq 0$ and $p_3^2=0$,
we get
\bea
\label{J_p3}
\left. \frac{}{} \!\! J(n;\nu_1,\nu_2,\nu_3)
\right|_{p_3^2=0}
= \mbox{i}^{1-n} \pi^{n/2} (p_1^2)^{n/2-\Sigma\nu_i} \;
\frac{\Gamma\left(\halfn-\nu_1-\nu_2\right)}
     {\Gamma(\nu_3) \Gamma\left(n-\sum\nu_i\right)}
\hspace{45mm}
\nn \\
\times \left\{
\frac{\Gamma\left(\halfn-\nu_1-\nu_3\right)
      \Gamma\left(\sum\nu_i-\halfn\right)}
     {\Gamma(\nu_2)} \;
_2F_1\left(\left. \begin{array}{c} \nu_1,\; \Sigma\nu_i-\halfn \\
                                  \nu_1+\nu_3-\halfn+1 \end{array}
\right| \frac{p_2^2}{p_1^2} \right)
\hspace{32mm}
\right.
\nn \\
\left.
+ \frac{\Gamma\left(\nu_1+\nu_3-\halfn\right)
        \Gamma\left(\halfn-\nu_3\right)}
       {\Gamma(\nu_1)} \;
\left(\frac{p_2^2}{p_1^2}\right)^{n/2-\nu_1-\nu_3} \;
_2F_1\left(\left. \begin{array}{c} \nu_2,\; \halfn-\nu_3 \\
                                  \halfn-\nu_1-\nu_3+1 \end{array}
\right| \frac{p_2^2}{p_1^2} \right)
\right\} , \;\;
\eea
where $_2F_1$ is the Gauss hypergeometric function. 
If we use the well-known formula of analytic continuation
of $_2F_1$ function from the argument $z$ to $1-z$ (with 
$z=p_2^2/p_1^2$), we reproduce the result presented
in Appendix~A of ref.~\cite{DKS} (namely, their $K$-integral
at $N=0$).
When $\nu_1=-s$ (where $s$ is a non-negative integer),
the second term in the braces of (\ref{J_p3}) vanishes, and we get
\bea
\label{J_p3_2}
\left. \frac{}{} \!\! J(n;-s,\nu_2,\nu_3)
\right|_{p_3^2=0}
= \mbox{i}^{1-n} \pi^{n/2} (p_1^2)^{n/2+s-\nu_2-\nu_3} \;
\hspace{70mm}
\nn \\
\times
\frac{\Gamma\left(\halfn+s-\nu_2\right) 
      \Gamma\left(\halfn+s-\nu_3\right)
      \Gamma\left(\nu_2+\nu_3-s-\halfn\right)}
     {\Gamma(\nu_2) \Gamma(\nu_3) \Gamma\left(n+s-\nu_2-\nu_3\right)} \;
_2F_1\left(\left. \begin{array}{c} -s,\;
                                  \nu_2\!+\!\nu_3\!-\!s\!-\!\halfn \\
                                  \nu_3-s-\halfn+1 \end{array}
\right| \frac{p_2^2}{p_1^2} \right) , \hspace{3mm}
\eea 
with a terminating $_2F_1$ series containing $(s+1)$ terms.

\section*{Appendix B: One-loop results for the ghost-gluon \\ 
          $\hspace*{39mm}$ vertex}
\setcounter{equation}{0}
\renewcommand{\thesection}{B}

At the zero-loop level, we have
\be
a^{(0)}(p^2,0,0)=a^{(0)}(0,p^2,0)=a^{(0)}(0,0,p^2)=1 ,
\ee
whereas all other ghost-gluon functions are equal to zero
at this order.

The diagrams contributing to the ghost-gluon vertex at the one-loop 
level are shown in Fig.~3 of ref.~\cite{DOT1}. The
general expressions listed in Appendix~D of \cite{DOT1} give
the following results in the on-shell limits of interest:
\bea
a^{(1)}(p^2,0,0)&=&
\frac{g^2\;\eta}{(4\pi)^{n/2}} \;
\frac{C_A}{16(n-4)} \; \kappa(p^2) \;
\nn \\
&& \times
\left[ 4(n-4)-2\xi (2n^2-9n+8)+\xi^2 (n-2) (n-4) \right],
\\
a^{(1)}(0,p^2,0)&=&
-\frac{g^2\;\eta}{(4\pi)^{n/2}} \;
\frac{C_A}{4(n-4)} \; \kappa(p^2) \;
\left[ 2+ \xi (n^2-6n+7) \right] ,
\\
a^{(1)}(0,0,p^2)&=&
\frac{g^2\;\eta}{(4\pi)^{n/2}} \;
\frac{C_A}{8(n-4)} \; \kappa(p^2) \;   
\left[ 2 (3n-8)-\xi (n-4) \right] ,
\\
b^{(1)}(p^2,0,0)&=&
\frac{g^2\;\eta}{(4\pi)^{n/2}} \;
\frac{C_A}{16(n-4)p^2} \; \kappa(p^2) \;
\nn \\
&& \times
\left[ 16 (n-4)+4\xi (2n^2-15n+30)-\xi^2 (n^2-8n+20)\right] ,
\\
b^{(1)}(0,p^2,0)&=&
\frac{g^2\;\eta}{(4\pi)^{n/2}} \;
\frac{C_A}{8(n-4)p^2} \; \kappa(p^2) \; 
\nn \\
&& \times
\left[ 8 (n-4)+4\xi (n^2-7n+13)-\xi^2 (n^2-8n+14)\right] ,
\\
b^{(1)}(0,0,p^2)&=&
\frac{g^2\;\eta}{(4\pi)^{n/2}} \;      
\frac{C_A}{4(n-4)p^2} \; \kappa(p^2) \;
\left[ 8+2\xi (2n-5)-\xi^2\right] ,
\\
c^{(1)}(p^2,0,0)&=&
-\frac{g^2\;\eta}{(4\pi)^{n/2}} \;     
\frac{C_A}{4(n-4)p^2} \; \kappa(p^2) \;
\xi \; (n-6+2\xi) ,
\\
c^{(1)}(0,p^2,0)&=&
\frac{g^2\;\eta}{(4\pi)^{n/2}} \;
\frac{C_A}{8(n-4)p^2} \; \kappa(p^2) \;
\nn \\   
&& \times
\left[4(n-6)+2\xi (n-2)(n-6)+\xi^2 (n^2-10n+20)\right] ,
\\
c^{(1)}(0,0,p^2)&=&
-\frac{g^2\;\eta}{(4\pi)^{n/2}} \;
\frac{C_A}{16(n-4)p^2} \; \kappa(p^2) \;
(n-6)\left[8+4\xi (n-2)+\xi^2 (n-4)\right] ,
\\
d^{(1)}(p^2,0,0)&=&
\frac{g^2\;\eta}{(4\pi)^{n/2}} \;
\frac{C_A}{16(n\!-\!4)p^2} \kappa(p^2)
\left[ 8 (n\!-\!6)\!-4\xi (5n\!-\!18)\!+\xi^2 (n^2\!-\!4n\!-\!4)\right] ,
\hspace*{9mm}
\\
d^{(1)}(0,p^2,0)&=&
-\frac{g^2\;\eta}{(4\pi)^{n/2}} \;     
\frac{C_A}{8(n-4)p^2} \; \kappa(p^2) \;
\nn \\
&& \times
\left[8 (n-6)+2\xi (2n^2-15n+32)-\xi^2 (n^2-8n+14)\right] ,
\\
d^{(1)}(0,0,p^2)&=&
\frac{g^2\;\eta}{(4\pi)^{n/2}} \;
\frac{C_A}{4(n-4)p^2} \; \kappa(p^2) \;
\left[ 2 (n-6)-2\xi (2n-5)+\xi^2\right] ,
\\
e^{(1)}(p^2,0,0)&=&
-\frac{g^2\;\eta}{(4\pi)^{n/2}} \;
\frac{C_A}{2(n-4)p^2} \; \kappa(p^2) \; 
\left[ 2+\xi (2n-9)+2\xi^2\right] ,
\\
e^{(1)}(0,p^2,0)&=& e^{(1)}(0,0,p^2)
\; = \; -\frac{g^2\;\eta}{(4\pi)^{n/2}} \;
\frac{C_A}{4 p^2} \; \kappa(p^2) \;
\left[ 2+\xi(n-3)\right] ,
\eea
where $\eta$ and $\kappa(p^2)$ are defined by eqs.~(\ref{eta})
and (\ref{kappa}), respectively.
These results are valid for arbitrary values of $n$ and $\xi$.
Note that there are no quark-loop contributions at the one-loop
level.

\section*{Appendix C: Two-point functions}
\setcounter{equation}{0}
\renewcommand{\thesection}{C}

For arbitrary values of $n$ and $\xi$, 
the results for the one-loop two-point functions (see eqs.~(\ref{gl_po})
and (\ref{gh_se})) are available elsewhere (see \cite{PT-book,Muta,DOT1}). 
For completeness, we also present them here: 
\be
J^{(1)}(p^2) =
\frac{g^2\;\eta}{(4\pi)^{n/2}} \;
\kappa(p^2)
\left\{ - \frac{C_A}{8}\left[ \frac{4(3n\!-\!2)}{n-1}
+4(2n\!-\!7)\xi-(n\!-\!4)\xi^2 \right]
+ 2 T \frac{n-2}{n-1} \right\},
\ee
\be
G^{(1)}(p^2) =  
\frac{g^2\;\eta}{(4\pi)^{n/2}} \;
\frac{C_A}{4} \; \kappa(p^2) \; \left[ 2 + (n-3) \xi \right] ,
\ee
where $\kappa(p^2)$ and $\eta$ are defined in eqs.~(\ref{kappa})
and (\ref{eta}), respectively.
The corresponding diagrams are shown in Fig.~2 of ref.~\cite{DOT1}.

Two-loop diagrams contributing to the gluon polarization operator
are shown in Fig.~3 of ref.~\cite{DOT2}.
Calculating their sum,
we get the following unrenormalized results \cite{DOT2}:
\bea
J^{(2,\xi)}(p^2) =
C_A^2 \frac{g^4 \; \eta^2}{(4\pi)^n}  (-p^2)^{-2\ep}
\left\{
\frac{1}{\ep^2}
\left(- \frac{25}{12}\!+\!\frac{5}{24}\xi\!+\!\frac{1}{4}\xi^2 \right)
\!+\!\frac{1}{\ep}
\left(- \frac{583}{72}\!+\! \frac{113}{144}\xi\!-\! \frac{19}{24}\xi^2
      \!+\!\frac{3}{8}\xi^3 \right)
\right.
\hspace*{-3.5mm}
\nn \\
\left.
- \frac{14311}{432}+ \zeta_3+ \frac{425}{864}\xi+ 2\xi \zeta_3
- \frac{71}{72}\xi^2+ \frac{9}{16}\xi^3 + \frac{1}{16} \xi^4
\right\}+{\cal{O}}(\ep),
\hspace{4mm}
\eea
\bea
J^{(2,q)}(p^2) =
C_A T \frac{g^4 \; \eta^2}{(4\pi)^n} \; (-p^2)^{-2\ep}
\left\{ 
\frac{1}{\ep^2}
\left(\frac{5}{3}- \frac{2}{3}\xi \right)
+\frac{1}{\ep}
\left(\frac{101}{18} + \frac{8}{9}\xi - \frac{2}{3}\xi^2 \right)
\right.
\nn \\
\left.
+ \frac{1961}{108}+ 8 \zeta_3+ \frac{142}{27}\xi- \frac{22}{9}\xi^2   
\right\}
\nn \\ 
+ C_F T \frac{g^4 \; \eta^2}{(4\pi)^n} \; (-p^2)^{-2\ep}
 \left\{ \frac{2}{\ep}+ \frac{55}{3} - 16 \zeta_3 \right\}
+{\cal{O}}(\ep),
\hspace{20mm}
\eea
where $C_A$, $T$ and $C_F$ are defined at the end of section~2.

The two-loop ghost self-energy diagrams are shown in 
Fig.~4 of ref.~\cite{DOT2}. Their sum yields the following
unrenormalized results:
\bea   
G^{(2,\xi)}(p^2)
= C_A^2 \frac{g^4 \; \eta^2}{(4\pi)^n} (-p^2)^{-2\ep} 
\left\{
\frac{1}{\ep^2}
\left( \frac{5}{4} +\frac{7}{16}\xi -\frac{1}{32}\xi^2 \right)
+\frac{1}{\ep}
\left( \frac{83}{16}+ \frac{7}{32}\xi \right)
\right.
\nn \\
\left.
+ \frac{599}{32}- \frac{3}{4} \zeta_3- \frac{9}{64}\xi
+ \frac{3}{8}\xi^2 - \frac{3}{16} \xi^2\zeta_3
\right\}+ {\cal{O}}(\ep).
\eea
\be
G^{(2,q)}(p^2) =
C_A T \frac{g^4 \; \eta^2}{(4\pi)^n} \; (-p^2)^{-2\ep}
 \left\{ - \frac{1}{2\ep^2} - \frac{7}{4\ep}
         - \frac{53}{8} \right\} + {\cal{O}}(\ep).
\ee

The discussion of renormalization, as well as the renormalized 
results for $J$ and $G$ can be found in section~8
of ref.~\cite{DOT2} (see also \cite{BL}, where these results in
the Feynman gauge are presented).

\section*{Appendix D: Boundary integrals}
\setcounter{equation}{0}
\renewcommand{\thesection}{D}

Below the case $p_1^2=p_2^2=0, \; p_3^2\equiv p^2$ is understood.
Using the integration-by-parts relations (\ref{ibp1})--(\ref{ibp8}),
we can reduce the integrals with six denominators to
integrals where some of the lines are shrunk, i.e.\ some of
the $\nu$'s vanish. Here we list explicit results for such 
``boundary'' integrals of interest.
Apart from the one-loop formulae listed in Appendix~A,
the following well-known (see, for example, in \cite{KL,Uss}) trick 
is useful, to combine two internal lines attached to a triple 
vertex with the remaining external line on shell:
\be
\label{trick}
\left.
\frac{1}{\left[(p-q)^2\right]^{\nu} \; \left( q^2 \right)^{\nu'}}
\right|_{p^2=0}
= \frac{\Gamma(\nu+\nu')}{\Gamma(\nu) \; \Gamma(\nu')}
\int\limits_0^1
\frac{\alpha^{\nu-1} (1-\alpha)^{\nu'-1} \; \mbox{d}\alpha}
     {\left[(q-\alpha p)^2\right]^{\nu+\nu'}} .
\ee
In fact, this is nothing but the Feynman parametrization of a product
of two propagators, where it is taken into account that
the difference of their momenta is light-like.

The diagrams formally corresponding to the boundary cases of
the integrals $K_3$ and $K_2$ (listed below) are drawn in Fig.~4
and Fig.~5, respectively.  
To distinguish an off-shell external line (corresponding to $p_3$)
from the on-shell ones (corresponding to $p_1$ and $p_2$),
the former is drawn as a double line, which can be associated with
the sum of $p_1$ and $p_2$. 
Note that when using the tensor decomposition in the sub-loops
we may get some integrals with positive $\nu_7$.
This is not a problem, because the relevant integrals
can be calculated for any $\nu_7$ (see below).
When the result is valid for an arbitrary $\nu_7$, the
corresponding line is solid; when it is valid only for
non-positive integer $\nu_7$, this is indicated by a dashed line,
as in Fig.~3b. 

The results (\ref{K_3_6}), (\ref{K_3_13}) and (\ref{K_2_13})
can be obtained by repeated use of the one-loop formulae
(see Appendix~A). To get other results, the trick (\ref{trick})
has been used. For the integrals (\ref{K_3_14}),
(\ref{K_2_1}), (\ref{K_2_2}) and (\ref{K_2_17}), this trick
was used for two pairs of propagators.
The notation $\Sigma\nu_i$ means $\sum_{i=1}^7\nu_i$,
i.e.\ the sum over all $\nu$'s involved (excluding those equal to 
zero). In particular, if $\nu_7=-s$ then 
$\Sigma\nu_i=\sum_{i=1}^6\nu_i-s$.

\subsection*{D.1: $K_3$ integrals}

The following boundary integrals can be expressed in terms
of one-term products of $\Gamma$ functions:
\bea
\label{K_3_6}
K_3(n;\nu_1,\nu_2,\nu_3,\nu_4,\nu_5,0,\nu_7)
= \mbox{i}^{2-2n} \pi^n (p^2)^{n-\Sigma\nu_i}\;
      \Gamma\!\left(\nu_1\!+\!\nu_3\!+\!\nu_5\!-\!\halfn\right) 
      \Gamma\!\left(\nu_2\!+\!\nu_4\!+\!\nu_7\!-\!\halfn\right)
\nn \\
\times\!
\frac{\Gamma\!\left(\halfn\!-\!\nu_1\!-\!\nu_5\right)
      \Gamma\!\left(\halfn\!-\!\nu_3\!-\!\nu_5\right)
      \Gamma\!\left(\halfn\!-\!\nu_2\!-\nu_7\right)
      \Gamma\!\left(\halfn\!-\!\nu_4\!-\nu_7\right)}
     {\Gamma(\nu_1) \Gamma(\nu_2) \Gamma(\nu_3) \Gamma(\nu_4)
      \Gamma\left(n-\nu_1-\nu_3-\nu_5\right)
      \Gamma\left(n-\nu_2-\nu_4-\nu_7\right)},
\hspace{5mm}
\eea
\bea
\label{K_3_14}
K_3(n;0,\nu_2,\nu_3,0,\nu_5,\nu_6,\nu_7)
= \mbox{i}^{2-2n} \pi^n (p^2)^{n-\Sigma\nu_i} \;
      \Gamma\!\left(\Sigma\nu_i-n\right)
\hspace{46mm}   
\nn \\
\times\!
\frac{\Gamma\!\left(\halfn\!-\!\nu_2\!-\!\nu_7\right)
      \Gamma\!\left(\halfn\!-\!\nu_3\!-\!\nu_5\right)
      \Gamma\!\left(\halfn\!-\!\nu_6\right)
      \Gamma\!\left(n\!-\!\nu_2\!-\!\nu_5\!-\!\nu_6\!-\!\nu_7\right)
      \Gamma\!\left(n\!-\!\nu_3\!-\!\nu_5\!-\!\nu_6\!-\!\nu_7\right)}
     {\Gamma(\nu_2) \Gamma(\nu_3) \Gamma(\nu_6)
      \Gamma\left(n-\nu_2-\nu_6-\nu_7\right)
      \Gamma\left(n-\nu_3-\nu_5-\nu_6\right)
      \Gamma\left({\textstyle{{3n}\over2}}-\Sigma\nu_i\right)},
\eea
\bea
\label{K_3_13}
K_3(n;0,\nu_2,0,\nu_4,\nu_5,\nu_6,\nu_7)
= \mbox{i}^{2-2n} \pi^n (p^2)^{n-\Sigma\nu_i} \;
      \Gamma\!\left(\Sigma\nu_i-n\right)
\hspace{43mm}
\nn \\
\times\!
\frac{\Gamma\!\left(\halfn\!-\!\nu_5\right)
      \Gamma\!\left(\halfn\!-\!\nu_6\right)
      \Gamma\!\left(\nu_5\!+\!\nu_6\!-\!\halfn\right)
      \Gamma\!\left(n\!-\!\nu_2\!-\!\nu_5\!-\!\nu_6\!-\!\nu_7\right)
      \Gamma\!\left(n\!-\!\nu_4\!-\!\nu_5\!-\!\nu_6\!-\!\nu_7\right)}
     {\Gamma(\nu_2) \Gamma(\nu_4) \Gamma(\nu_5) \Gamma(\nu_6)
      \Gamma\left(n-\nu_5-\nu_6\right)
      \Gamma\left({\textstyle{{3n}\over2}}-\Sigma\nu_i\right)}.
\eea

In the following formula we assume that $\nu_7=-s$ is a non-positive
integer:
\bea
\label{K_3_45}
K_3(n;\nu_1,\nu_2,\nu_3,0,0,\nu_6,-s)
= \mbox{i}^{2-2n} \pi^n (p^2)^{n-\Sigma\nu_i} \;
      \Gamma\!\left(\Sigma\nu_i-n\right)
\hspace{37mm}
\nn \\
\times\!
\frac{\Gamma\!\left(\halfn\!+\!s\!-\!\nu_2\!\right)
      \Gamma\!\left(\halfn\!-\!\nu_3\!\right)
      \Gamma\!\left(\halfn\!+\!s\!-\!\nu_6\!\right)
      \Gamma\!\left(n\!+\!s\!-\!\nu_1\!-\!\nu_2\!-\!\nu_6\right)}
     {\Gamma(\nu_2) \Gamma(\nu_3) \Gamma(\nu_6)
      \Gamma\left({\textstyle{{3n}\over2}} - \Sigma\nu_i \right)
      \Gamma\left(n\!+\!s\!-\!\nu_2\!-\!\nu_6\right)}
\hspace{25mm}
\nn \\
\times\!
\frac{\Gamma\left(\nu_2+\nu_6-s-\halfn\right)}
     {\Gamma\left(\nu_1+\nu_2+\nu_6-s-\halfn\right)} \;
_3F_2\left( \left.
\begin{array}{c} -s,\; \halfn-\nu_3, \; \nu_2+\nu_6-s-\halfn \\
                 \nu_6\!-\!s\!-\!\halfn\!+\!1, \;
                 \nu_1\!+\!\nu_2\!+\!\nu_6\!-\!s\!-\!\halfn \end{array}
\right| 1 \right) .
\eea
Here, $_3F_2$ denotes a generalized hypergeometric series. In fact,
we have a terminating $_3F_2$ series
of unit argument, since one of the upper parameters is
equal to $-s$. This may be considered just as a compact representation
of a finite sum containing $(s+1)$ terms, each term being
a product of $\Gamma$ functions. 

Using the results for the boundary integrals with $\nu_7>0$,
one can also calculate the planar forward-scattering double-box 
diagram (see Fig.~3b),
\be
\label{fsdb}
K_3(n;1,1,1,1,1,1,1) =
-\pi^{4-2\ep}(-p^2)^{-3-2\ep} \eta^2 
\left[ \frac{1}{\ep^3} \!-\! \frac{5}{\ep^2} \!+\! \frac{16}{\ep}
\!+\!\frac{\pi^2}{\ep}\!-\!44\!-\!\pi^2\!+\!24\zeta_3 
\!+\! {\cal{O}}(\ep) \right],
\ee
and similar integrals with higher powers of the propagators.

\subsection*{D.2: $K_2$ integrals}

In the following two formulae, we also assume that $\nu_7=-s$ is 
a non-positive integer:
\bea
\label{K_2_1}
K_2(n;0,\nu_2,\nu_3,\nu_4,\nu_5,\nu_6,-s)
= \mbox{i}^{2-2n} \pi^n (p^2)^{n-\Sigma\nu_i} \;
      \Gamma\!\left(\Sigma\nu_i-n\right)
\hspace{47mm}
\nn \\
\times\!
\frac{\Gamma\!\left(\halfn\!-\!\nu_2\!-\!\nu_4\right)
      \Gamma\!\left(\halfn\!-\!\nu_3\!-\!\nu_5\right)
      \Gamma\!\left(\halfn\!-\!\nu_6\!\right)
      \Gamma\!\left(n\!-\!\nu_2\!-\!\nu_3\!-\!\nu_4\!-\!\nu_6\right)
      \Gamma\!\left(n\!+\!s\!-\!\nu_3\!-\!\nu_4\!-\!\nu_5\!-\!\nu_6\right)}
     {\Gamma(\nu_2) \Gamma(\nu_5) \Gamma(\nu_6)
      \Gamma\left({\textstyle{{3n}\over2}} - \Sigma\nu_i \right)
      \Gamma\left(n\!-\!\nu_2\!-\!\nu_4\!-\!\nu_6\right)
      \Gamma\left(n\!-\!\nu_3\!-\!\nu_5\!-\!\nu_6\right)}
\nn \\
\times\!
\frac{\Gamma\left(\nu_3+\nu_5+\nu_6-\halfn\right)}   
     {\Gamma\left(\nu_3+\nu_5+\nu_6-s-\halfn\right)} \;
_3F_2\left( \left.
\begin{array}{c} -s,\; \nu_3, \; \halfn-\nu_2-\nu_4 \\
                 n\!-\!\nu_2\!-\!\nu_4\!-\!\nu_6, \;
                 \nu_3\!+\!\nu_5\!+\!\nu_6\!-\!s\!-\!\halfn \end{array}
\right| 1 \right) ,
\hspace{10mm}
\eea
\bea  
\label{K_2_2}
K_2(n;\nu_1,0,\nu_3,\nu_4,\nu_5,\nu_6,-s)
= \mbox{i}^{2-2n} \pi^n (p^2)^{n-\Sigma\nu_i} \;
      \Gamma\left(\Sigma\nu_i-n\right)
\hspace{40mm}
\nn \\
\times\!
\frac{\Gamma\!\left(\halfn\!-\!\nu_4\!+\!s\right)
      \Gamma\!\left(\halfn\!-\!\nu_6\!+\!s\right)
      \Gamma\!\left(n\!-\!\nu_1-\!\nu_3\!-\!\nu_4\!-\!\nu_6+\!s\right)
      \Gamma\!\left(n\!-\!\nu_3\!-\!\nu_4-\!\nu_5\!-\!\nu_6+\!s\right)}
     {\Gamma(\nu_1) \Gamma(\nu_4) \Gamma(\nu_5) \Gamma(\nu_6)
      \Gamma\left(n\!-\!\nu_4\!-\!\nu_6\!+\!s\right)   
      \Gamma\left({\textstyle{{3n}\over2}}\!
           -\!\nu_1\!-\!\nu_3\!-\!\nu_4\!-\!\nu_5\!-\!\nu_6+\!s\right)}
\nn \\
\times
\Gamma\!\left(\nu_4\!+\!\nu_6\!-\!s\!-\!\halfn\right) \;\;
_3F_2\left( \left.
\begin{array}{c} -s,\; 1-\nu_5, \; \nu_4+\nu_6-s-\halfn \\
                 \nu_6\!-\!s\!-\!\halfn\!+\!1, \;
             \nu_1\!+\!\nu_3\!+\!\nu_4\!+\!\nu_6\!-\!s\!-\!n\!+\!1
\end{array}
\right| 1 \right) .
\hspace{5mm}
\eea
Here, we also have terminating $_3F_2$ series of unit argument,
containing $(s+1)$ terms. However, in some cases they may 
contain less terms. For example, for an integer $\nu_5>0$ 
the number of terms in $_3F_2$ from (\ref{K_2_2}) is 
$\min(s+1,\nu_5)$. In particular, when $s=0$ or $\nu_5=1$
we get just one term.

When $\nu_3=0$, eq.~(\ref{K_2_1}) also contains just one term, 
since one of the upper parameters in $_3F_2$ vanishes, 
so that $_3F_2=1$. Moreover, the corresponding result can be
extended to an arbitrary value of $\nu_7$:
\bea
\label{K_2_13}
K_2(n;0,\nu_2,0,\nu_4,\nu_5,\nu_6,\nu_7)
= \mbox{i}^{2-2n} \pi^n (p^2)^{n-\Sigma\nu_i} \;
      \Gamma\left(\Sigma\nu_i-n\right)
\hspace{38mm}
\nn \\
\times\!
\frac{\Gamma\!\left(\halfn\!-\!\nu_5\right)
      \Gamma\!\left(\halfn\!-\!\nu_6\right)
      \Gamma\!\left(\halfn\!-\!\nu_2\!-\!\nu_4\right)
      \Gamma\!\left(\nu_5\!+\!\nu_6\!-\!\halfn\right)
      \Gamma\!\left(n\!-\!\nu_4\!-\!\nu_5\!-\!\nu_6\!-\!\nu_7\right)}
     {\Gamma(\nu_2) \Gamma(\nu_5) \Gamma(\nu_6)
      \Gamma\!\left(\nu_5\!+\!\nu_6\!+\!\nu_7\!-\!\halfn\right)
      \Gamma\left(n-\nu_5-\nu_6\right)
      \Gamma\left({\textstyle{{3n}\over2}}-\Sigma\nu_i\right)}.
\eea

An important special case of eq.~(\ref{K_2_1}) is $\nu_7=0$ ($s=0$).
In this case, the $_3F_2$ function is equal to 1 (since one
of the upper parameters is zero) and we get\footnote{This diagram
has been considered in ref.~\cite{SuSch2}, using the method
of negative-dimensional integration. 
We note that in their result (11) (or in the definition (5))
the parameters $m$ and $n$ (corresponding, up to a sign, to
$\nu_4$ and $\nu_3$ in our eq.~(\ref{K_2_17})) should be interchanged.
Also, $(-\pi)^D$ should read $\pi^D$ (they use Euclidean metric). 
We are grateful to the authors of \cite{SuSch2} for confirming
these misprints.
The other diagrams considered in \cite{SuSch2} (see also 
\cite{SuSch}) correspond
to integrals that can be obtained by repeated use of
one-loop formulae. The results are given in eqs. (18), (21) and (23) 
of \cite{SuSch2}, and they correspond to the special cases of
our eqs.~(\ref{K_2_2}) (at $s=0$), (\ref{K_3_6}) (at $\nu_7=0$) 
and (\ref{K_3_13}) (at $\nu_7=0$), respectively.}
\bea
\label{K_2_17}
K_2(n;0,\nu_2,\nu_3,\nu_4,\nu_5,\nu_6,0)
= \mbox{i}^{2-2n} \pi^n (p^2)^{n-\Sigma\nu_i} \;
      \Gamma\left(\Sigma\nu_i-n\right)
\hspace{47mm}
\nn \\
\times\!
\frac{\Gamma\!\left(\halfn\!-\!\nu_2\!-\!\nu_4\right)
      \Gamma\!\left(\halfn\!-\!\nu_3\!-\!\nu_5\right)
      \Gamma\!\left(\halfn\!-\!\nu_6\right)
      \Gamma\!\left(n\!-\!\nu_2\!-\!\nu_3\!-\!\nu_4\!-\!\nu_6\right)
      \Gamma\!\left(n\!-\!\nu_3\!-\!\nu_4\!-\!\nu_5\!-\!\nu_6\right)}
     {\Gamma(\nu_2) \Gamma(\nu_5) \Gamma(\nu_6)
      \Gamma\left(n\!-\!\nu_2\!-\!\nu_4\!-\!\nu_6\right)
      \Gamma\left(n\!-\!\nu_3\!-\!\nu_5\!-\!\nu_6\right)
      \Gamma\left({\textstyle{{3n}\over2}}-\Sigma\nu_i\right)}. 
\nonumber \\
\eea
This result corresponds to the last diagram shown in Fig.~5.

\section*{Appendix E: Two-loop results for the ghost-gluon \\ 
          $\hspace*{39mm}$ vertex}
\setcounter{equation}{0}
\renewcommand{\thesection}{E}

Here we present the unrenormalized expressions for two-loop 
contributions to the scalar
functions occurring in the ghost-gluon vertex (\ref{BC-ghg}),
in all on-shell limits of interest. To calculate these functions,
the same algorithms (and the same {\sf REDUCE} program) as
for the three-gluon vertex have been employed.
The two-loop diagrams contributing to the ghost-gluon vertex
are shown in Fig.~2 of ref.~\cite{DOT2}.
Their renormalization is similar to that of the three-gluon
vertex (see section~8 and Appendix~B of ref.~\cite{DOT2});
the renormalization factor $\widetilde{Z}_1$ should be used.

\subsection*{E.1: Non-zero gluon momentum squared}

\bea
a^{(2,\xi)}(p^2,0,0) & \!=\! &
C_A^2 \; \frac{g^4\;\eta^2}{(4\pi)^n}
(-p^2)^{-2\ep}
\left\{
\frac{\xi}{\ep^4} \left( \!- \frac{1}{32} \!+\! \frac{1}{128} \xi \!\right)  
\!+ \frac{1}{\ep^3} \left( \frac{5}{96} \!+\! \frac{17}{192} \xi
\!-\! \frac{21}{128} \xi^2 \!+\! \frac{1}{128} \xi^3 \!\right)
\right.
\nn \\
&& +  \frac{1}{\ep^2} \left(
\frac{19}{72} \!-\! \frac{5}{96} \pi^2 \!-\! \frac{25}{288} \xi 
\!-\! \frac{3}{128} \pi^2\xi \!+\! \frac{73}{384} \xi^2
\!+\! \frac{1}{384} \pi^2 \xi^2 \!-\! \frac{19}{128} \xi^3
\!+\! \frac{1}{768} \pi^2 \xi^3 \right)
\nn \\
&& +  \frac{1}{\ep} \left(
\frac{2141}{864}- \frac{101}{288} \pi^2 - \frac{5}{16} \zeta_3
+ \frac{125}{1728} \xi- \frac{45}{64} \zeta_3 \xi
- \frac{1}{72} \xi^2 - \frac{5}{384}  \pi^2\xi^2  
\right.
\nn \\
&& \left.
- \frac{1}{32} \zeta_3 \xi^2
 - \frac{1}{8} \xi^3+ \frac{7}{768} \pi^2 \xi^3 + \frac{1}{128} \zeta_3 \xi^3
- \frac{1}{32} \xi^4 \right)
\nn \\
&& + \frac{57245}{5184} - \frac{895}{864} \pi^2
- \frac{179}{48} \zeta_3 - \frac{1}{64} \pi^4
- \frac{9139}{10368} \xi + \frac{3}{32} \pi^2 \xi
- \frac{81}{32} \zeta_3 \xi 
\nn \\
&& - \frac{21}{1280} \pi^4 \xi
+ \frac{587}{1728} \xi^2 - \frac{1}{384} \pi^2 \xi^2
- \frac{7}{16} \zeta_3 \xi^2
 - \frac{23}{64} \xi^3 + \frac{19}{768} \pi^2 \xi^3
\nn \\
&& \left.
+ \frac{7}{128} \zeta_3 \xi^3 + \frac{1}{2560} \pi^4 \xi^3
- \frac{3}{32} \xi^4 
\right\}
+ {\cal{O}}(\ep),
\eea
\bea
p^2 b^{(2,\xi)}(p^2,0,0) & \!\!=\!\! &
C_A^2 \; \frac{g^4\;\eta^2}{(4\pi)^n}
(-p^2)^{-2\ep}
\left\{
\frac{\xi}{\ep^4}
\left( \frac{3}{64} \!-\! \frac{3}{128} \xi \right)  
+ \frac{1}{\ep^3} 
\left( - \frac{1}{3}\! +\! \frac{19}{96} \xi\! + \frac{1}{24} \xi^2 
\!- \frac{13}{128} \xi^3 \right)
\right.
\nn \\
&& +  \frac{1}{\ep^2} 
\left( \frac{17}{72} + \frac{1}{24} \pi^2 + \frac{17}{18} \xi
    + \frac{1}{192} \pi^2 \xi - \frac{83}{576} \xi^2
   - \frac{7}{384} \pi^2 \xi^2 - \frac{5}{32} \xi^3
\right.
\nn \\   
&& \left.
   + \frac{1}{384} \pi^2 \xi^3 - \frac{1}{32} \xi^4 \right) 
\nn \\
&& + \frac{1}{\ep} 
\left(  \frac{779}{432}  + \frac{31}{144} \pi^2 + \frac{1}{4} \zeta_3
   + \frac{1093}{864} \xi + \frac{1}{96} \pi^2 \xi  
   + \frac{1}{2} \zeta_3 \xi 
   + \frac{331}{864} \xi^2 - \frac{1}{64} \pi^2 \xi^2
\right.
\nn \\
&& \left.
   - \frac{11}{32} \zeta_3 \xi^2
   - \frac{67}{128} \xi^3 + \frac{1}{128} \pi^2 \xi^3
   + \frac{1}{64} \zeta_3 \xi^3 - \frac{1}{8} \xi^4 \right)
\nn \\
&& + \frac{24959}{2592} + \frac{293}{432} \pi^2 \!+ \frac{25}{24} \zeta_3
+ \frac{1}{80} \pi^4 
\!+ \frac{5371}{5184} \xi - \frac{11}{96} \pi^2 \xi
+ \frac{35}{16} \zeta_3 \xi \!+ \frac{3}{320} \pi^4 \xi
\nn \\
&& + \frac{5549}{2592} \xi^2 + \frac{7}{192} \pi^2 \xi^2
- \frac{7}{4} \zeta_3 \xi^2 - \frac{3}{320} \pi^4 \xi^2
- \frac{11}{8} \xi^3 + \frac{1}{48} \pi^2 \xi^3
+ \frac{3}{16} \zeta_3 \xi^3
\nn \\
&& \left.
+ \frac{1}{1280} \pi^4 \xi^3
- \frac{13}{32} \xi^4
\right\}
+ {\cal{O}}(\ep),
\eea
\bea   
p^2 c^{(2,\xi)}(p^2,0,0) & \!\!=\!\! &
C_A^2 \; \frac{g^4\;\eta^2}{(4\pi)^n}
(-p^2)^{-2\ep}
\left\{
\frac{\xi}{\ep^4}
\left( \frac{1}{64} - \frac{3}{128} \xi + \frac{1}{256} \xi^2 \right)
\right.
\nn \\
&& + \frac{1}{\ep^3} 
\left( - \frac{25}{48} + \frac{7}{32} \xi - \frac{7}{384} \xi^2
     - \frac{45}{256} \xi^3 \right)
\nn \\
&& + \frac{1}{\ep^2} 
\left( - \frac{163}{72} - \frac{1}{24} \pi^2 
   + \frac{5}{4} \xi - \frac{1}{32} \pi^2 \xi
   - \frac{5}{72} \xi^2 - \frac{1}{128} \pi^2 \xi^2
   - \frac{85}{256} \xi^3 
\right.  
\nn \\   
&& \left.
+ \frac{3}{256} \pi^2 \xi^3
   - \frac{1}{16} \xi^4  \right)
\nn \\
&& + \frac{1}{\ep} 
\left( - \frac{2407}{432} - \frac{3}{8} \pi^2 
  - \frac{1}{4} \zeta_3
   + \frac{409}{96} \xi - \frac{1}{8} \pi^2 \xi
   - \frac{9}{32} \zeta_3 \xi
   - \frac{1331}{1728} \xi^2 
\right.
\nn \\
&& \left.
+ \frac{9}{128} \pi^2 \xi^2
   - \frac{15}{32} \zeta_3 \xi^2
   - \frac{77}{128} \xi^3 + \frac{7}{256} \pi^2 \xi^3
   + \frac{3}{64} \zeta_3 \xi^3 - \frac{1}{4} \xi^4 \right)
\nn \\
&& - \frac{26449}{2592} - \frac{29}{16} \pi^2
- \frac{1}{4} \zeta_3 - \frac{1}{80} \pi^4
+ \frac{4651}{576} \xi - \frac{11}{48} \pi^2 \xi
- \frac{5}{16} \zeta_3 \xi  - \frac{7}{640} \pi^4 \xi
\nn \\
&& - \frac{4471}{5184} \xi^2 + \frac{73}{192} \pi^2 \xi^2
 - \frac{87}{32} \zeta_3 \xi^2 - \frac{3}{320} \pi^4 \xi^2
- \frac{17}{8} \xi^3 + \frac{79}{768} \pi^2 \xi^3
+ \frac{3}{8} \zeta_3 \xi^3 
\nn \\
&& \left.
+ \frac{1}{320} \pi^4 \xi^3 - \frac{3}{4} \xi^4
\right\}
+ {\cal{O}}(\ep),
\eea
\bea   
p^2 d^{(2,\xi)}(p^2,0,0) & \!\!=\!\! &
C_A^2 \; \frac{g^4\;\eta^2}{(4\pi)^n}
(-p^2)^{-2\ep}
\left\{
- \frac{1}{\ep^4}
\left( \frac{3}{32} + \frac{1}{64} \xi + \frac{1}{128} \xi^2
\right) 
\right.
\nn \\
&& + \frac{1}{\ep^3}
\left( \frac{5}{48}  + \frac{5}{16} \xi
    - \frac{55}{192} \xi^2 - \frac{11}{128} \xi^3 \right) 
\nn \\
&& + \frac{1}{\ep^2} 
\left( \frac{119}{72} - \frac{3}{32} \pi^2 
   + \frac{17}{24} \xi  - \frac{1}{32} \pi^2 \xi
   + \frac{263}{1152} \xi^2 - \frac{7}{384} \pi^2 \xi^2
   - \frac{29}{64} \xi^3  
\right.  
\nn \\   
&& \left.
   + \frac{1}{192} \pi^2 \xi^3
   - \frac{1}{32} \xi^4  \right)
\nn \\
&& + \frac{1}{\ep} 
\left( \frac{248}{27} - \frac{1}{2} \pi^2 - \frac{3}{4} \zeta_3
   + \frac{31}{18} \xi + \frac{1}{32} \pi^2 \xi
   - \frac{27}{32} \zeta_3 \xi
   + \frac{307}{864} \xi^2  - \frac{1}{16} \pi^2 \xi^2 
\right.
\nn \\
&& \left.
   - \frac{7}{16} \zeta_3 \xi^2
   - \frac{99}{128} \xi^3 + \frac{5}{192} \pi^2 \xi^3
   + \frac{1}{32} \zeta_3 \xi^3 - \frac{3}{16} \xi^4 \right)  
\nn \\
&& + \frac{24947}{648} - \frac{103}{72} \pi^2
- \frac{17}{2} \zeta_3 - \frac{1}{32} \pi^4
+ \frac{523}{432} \xi + \frac{11}{96} \pi^2 \xi
- \frac{25}{8} \zeta_3 \xi - \frac{13}{640} \pi^4 \xi 
\nn \\
&& + \frac{7553}{2592} \xi^2 - \frac{13}{384} \pi^2 \xi^2
- \frac{11}{4} \zeta_3 \xi^2 - \frac{7}{640} \pi^4 \xi^2
- \frac{67}{32} \xi^3 + \frac{9}{128} \pi^2 \xi^3
+ \frac{19}{64} \zeta_3 \xi^3
\nn \\
&& \left.
+ \frac{1}{640} \pi^4 \xi^3 - \frac{19}{32} \xi^4
\right\}
+ {\cal{O}}(\ep),
\eea
\bea   
p^2 e^{(2,\xi)}(p^2,0,0) & \!\!=\!\! &
C_A^2 \; \frac{g^4\;\eta^2}{(4\pi)^n}
(-p^2)^{-2\ep}
\left\{
\frac{1}{\ep^4}
\left( - \frac{3}{32} + \frac{1}{64} \xi - \frac{3}{64} \xi^2
   + \frac{1}{128} \xi^3 \right) 
\right.
\nn \\
&& + \frac{1}{\ep^3}
\left(  - \frac{11}{16} + \frac{23}{32} \xi
   - \frac{11}{96} \xi^2  - \frac{45}{128} \xi^3 \right)
\nn \\
&& + \frac{1}{\ep^2}
\left(  - \frac{19}{12}  - \frac{7}{48} \pi^2 
    + \frac{93}{32} \xi - \frac{1}{16} \pi^2 \xi
    - \frac{29}{144} \xi^2 - \frac{1}{32} \pi^2 \xi^2
    - \frac{85}{128} \xi^3 
\right.  
\nn \\   
&& \left.
    + \frac{3}{128} \pi^2 \xi^3
    - \frac{1}{8} \xi^4 \right)
\nn \\  
&& + \frac{1}{\ep}
\left( \frac{35}{288} - \frac{67}{72} \pi^2 - \frac{17}{16} \zeta_3 
    + \frac{1681}{192} \xi - \frac{17}{96} \pi^2 \xi
    - \frac{27}{32} \zeta_3 \xi
    - \frac{1439}{864} \xi^2 
\right.
\nn \\
&& \left.
    + \frac{7}{64} \pi^2 \xi^2
    - \frac{33}{32} \zeta_3 \xi^2
    - \frac{77}{64} \xi^3 + \frac{7}{128} \pi^2 \xi^3
    + \frac{3}{32} \zeta_3 \xi^3 - \frac{1}{2} \xi^4 \right)
\nn \\   
&& + \frac{33431}{1728} - \frac{1795}{432} \pi^2
- \frac{125}{24} \zeta_3 - \frac{3}{64} \pi^4
+ \frac{19117}{1152} \xi - \frac{5}{16} \pi^2 \xi
- \frac{9}{4} \zeta_3 \xi 
\nn \\
&& - \frac{17}{640} \pi^4 \xi
- \frac{409}{648} \xi^2 + \frac{55}{96} \pi^2 \xi^2
- \frac{171}{32} \zeta_3 \xi^2 - \frac{3}{128} \pi^4 \xi^2
- \frac{17}{4} \xi^3 + \frac{79}{384} \pi^2 \xi^3
\nn \\
&& \left.
+ \frac{3}{4} \zeta_3 \xi^3 + \frac{1}{160} \pi^4 \xi^3
- \frac{3}{2} \xi^4
\right\}
+ {\cal{O}}(\ep),
\eea
\bea
a^{(2,q)}(p^2,0,0) & \!=\! &
C_A T \; \frac{g^4\;\eta^2}{(4\pi)^n}
(-p^2)^{-2\ep}
\left\{
\frac{1}{\ep^3}
\left( - \frac{1}{6} + \frac{1}{6} \xi  \right) 
+ \frac{1}{\ep^2}
\left( - \frac{7}{36} + \frac{1}{36} \xi + \frac{1}{6} \xi^2 \right)
\right.
\nn \\
&& + \frac{1}{\ep}
\left( - \frac{235}{216} + \frac{5}{36} \pi^2
    + \frac{11}{108} \xi + \frac{4}{9} \xi^2 \right)
\nn \\
&& \left. 
- \frac{4963}{1296} + \frac{77}{216} \pi^2 + \frac{11}{6} \zeta_3 
+ \frac{157}{324} \xi + \zeta_3 \xi + \frac{61}{54} \xi^2 
\right\}
+{\cal{O}}(\ep),
\eea
\bea
p^2 b^{(2,q)}(p^2,0,0) & \!\!=\!\! &
C_A T \; \frac{g^4\;\eta^2}{(4\pi)^n}   
(-p^2)^{-2\ep}
\left\{
\frac{1}{\ep^3}
\left( \frac{1}{6} - \frac{1}{3} \xi + \frac{1}{6} \xi^2 \right) 
+ \frac{1}{\ep^2}
\left( - \frac{1}{18}  - \frac{11}{9} \xi + \frac{11}{18} \xi^2
\right)
\right.
\nn \\
&&   + \frac{1}{\ep}
\left( \frac{23}{108} - \frac{2}{9} \pi^2 
    - \frac{103}{27} \xi + \frac{103}{54} \xi^2 \right)
\nn \\
&& \left.
- \frac{313}{648} - \frac{67}{108} \pi^2 
- \frac{7}{3} \zeta_3
- \frac{827}{81} \xi + \frac{827}{162} \xi^2
\right\}
+{\cal{O}}(\ep),
\eea   
\bea
p^2 c^{(2,q)}(p^2,0,0) & \!\!=\!\! &
C_A T \; \frac{g^4\;\eta^2}{(4\pi)^n}
(-p^2)^{-2\ep}
\left\{
\frac{1}{\ep^3}
\left( \frac{1}{6} - \frac{1}{2} \xi + \frac{1}{3} \xi^2 \right)
+  \frac{1}{\ep^2}
\left( \frac{17}{18} - 2 \xi + \frac{11}{9} \xi^2 \right) 
\right.
\nn \\ 
&& + \frac{1}{\ep}
\left( \frac{239}{108} + \frac{1}{6} \pi^2
    - \frac{16}{3} \xi + \frac{94}{27} \xi^2  \right)
\nn \\
&& \left.
+ \frac{2387}{648} + \frac{11}{12} \pi^2
- \frac{116}{9} \xi + \zeta_3 \xi  + \frac{728}{81} \xi^2 
\right\}
+{\cal{O}}(\ep),
\eea
\bea
p^2 d^{(2,q)}(p^2,0,0) & \!\!=\!\! &
-C_A T \; \frac{g^4\;\eta^2}{(4\pi)^n}
(-p^2)^{-2\ep}
\left\{
\frac{1}{\ep^3}
\left( \frac{1}{3} - \frac{1}{6} \xi^2 \right) 
+ \frac{1}{\ep^2}
\left( \frac{8}{9}  + \frac{7}{6} \xi - \frac{17}{18} \xi^2
 \right) 
\right. 
\nn \\
&& + \frac{1}{\ep}
\left( \frac{88}{27} + \frac{65}{18} \xi
- \frac{151}{54} \xi^2 \right) 
\nn \\
&& \left. 
+ \frac{1945}{162} + \frac{1}{18} \pi^2 - 2 \zeta_3
+ \frac{499}{54} \xi - 2 \zeta_3 \xi - \frac{1193}{162} \xi^2
\right\}
+{\cal{O}}(\ep),
\eea
\bea
p^2 e^{(2,q)}(p^2,0,0) & \!\!=\!\! &
C_A T \; \frac{g^4\;\eta^2}{(4\pi)^n}
(-p^2)^{-2\ep}
\left\{
\frac{\xi}{\ep^3}
\left( - 1 + \frac{2}{3} \xi \right) 
+ \frac{1}{\ep^2} 
\left( \frac{5}{12} - 4 \xi + \frac{22}{9} \xi^2 \right)
\right.
\nn \\
&& + \frac{1}{\ep}
\left( - \frac{73}{72} + \frac{5}{18} \pi^2
    - \frac{32}{3} \xi + \frac{188}{27} \xi^2 \right)
\nn \\ 
&& \left.
- \frac{4801}{432} + \frac{44}{27} \pi^2 + \frac{5}{3} \zeta_3 
- \frac{232}{9} \xi + 2 \zeta_3 \xi  + \frac{1456}{81} \xi^2
\right\}   
+{\cal{O}}(\ep).
\eea

\subsection*{E.2: Non-zero in-ghost momentum squared}

\bea
a^{(2,\xi)}(0,p^2,0) & \!=\! &
C_A^2 \; \frac{g^4\;\eta^2}{(4\pi)^n}
(-p^2)^{-2\ep}
\left\{
\frac{1}{\ep^4}
\left(- \frac{3}{64} + \frac{3}{128} \xi \right)
+ \frac{1}{\ep^3}
\left( \frac{5}{96} - \frac{21}{128} \xi + \frac{19}{256} \xi^2
\right)
\right.
\nn \\
&& + \frac{1}{\ep^2}
\left( \frac{31}{72} - \frac{13}{192} \pi^2
    - \frac{93}{64} \xi + \frac{7}{384} \pi^2 \xi
    + \frac{67}{128} \xi^2 - \frac{1}{384} \pi^2 \xi^2 \right)
\nn \\
&& +  \frac{1}{\ep}
\left( \frac{355}{108} - \frac{11}{36} \pi^2 - \frac{1}{2} \zeta_3
- \frac{305}{64} \xi + \frac{11}{128} \pi^2 \xi
+ \frac{5}{32} \zeta_3 \xi
+ \frac{187}{128} \xi^2 
\right.
\nn \\
&& \left.
- \frac{11}{768} \pi^2 \xi^2
- \frac{1}{64} \zeta_3 \xi^2 \right)
\nn \\
&& + \frac{37027}{2592} - \frac{791}{864} \pi^2
- \frac{175}{48} \zeta_3 - \frac{7}{320} \pi^4
- \frac{263}{16} \xi + \frac{23}{96} \pi^2 \xi
+ \frac{33}{16} \zeta_3 \xi + \frac{1}{160} \pi^4 \xi
\nn \\
&& \left.
 + \frac{541}{128} \xi^2 - \frac{3}{64} \pi^2 \xi^2
- \frac{1}{16} \zeta_3 \xi^2 - \frac{1}{1280} \pi^4 \xi^2
\right\}
+ {\cal{O}}(\ep),
\eea
\bea   
p^2 b^{(2,\xi)}(0,p^2,0) & \!\!=\!\! &
C_A^2 \; \frac{g^4\;\eta^2}{(4\pi)^n}
(-p^2)^{-2\ep}
\left\{
\frac{\xi}{\ep^4}
\left(  - \frac{1}{64} + \frac{1}{128} \xi + \frac{1}{128} \xi^2
\right) 
\right.
\nn \\
&& + \frac{1}{\ep^3} 
\left( - \frac{1}{6} - \frac{73}{192} \xi
   + \frac{11}{128} \xi^2 + \frac{7}{64} \xi^3 \right) 
\nn \\
&& + \frac{1}{\ep^2} 
\left( \frac{407}{144} \!-\! \frac{5}{48} \pi^2 
   \!-\! \frac{323}{144} \xi \!+\! \frac{1}{32} \pi^2 \xi
   \!-\! \frac{1}{128} \xi^2 \!+\! \frac{1}{96} \pi^2 \xi^2
   \!+ \frac{85}{256} \xi^3 \!+\! \frac{1}{384} \pi^2 \xi^3\!
\right)
\nn \\
&&  + \frac{1}{\ep} 
\left(  \frac{13391}{864}  - \frac{31}{48} \pi^2
   - \frac{5}{8} \zeta_3
   - \frac{14383}{1728} \xi - \frac{1}{64} \pi^2 \xi
   + \frac{9}{32} \zeta_3 \xi
   + \frac{75}{64} \xi^2
\right.
\nn \\
&& \left.
   - \frac{11}{384} \pi^2 \xi^2
   + \frac{25}{64} \zeta_3 \xi^2
   + \frac{101}{128} \xi^3 + \frac{1}{96} \pi^2 \xi^3
   - \frac{1}{32} \zeta_3 \xi^3 \right)
\nn \\
&&  + \frac{315815}{5184} - \frac{17}{9} \pi^2 
\!- \frac{47}{8} \zeta_3 \!- \frac{1}{32} \pi^4
\!- \frac{376861}{10368} \xi \!+ \frac{31}{48} \pi^2 \xi
+ \frac{35}{16} \zeta_3 \xi \!+ \frac{7}{640} \pi^4 \xi
\nn \\   
&& \left.
+ \frac{131}{32} \xi^2 \!- \frac{25}{384} \pi^2 \xi^2
+ \frac{13}{16} \zeta_3 \xi^2 \!+ \frac{11}{1280} \pi^4 \xi^2
+ \frac{17}{8} \xi^3 \!+ \frac{23}{768} \pi^2 \xi^3
- \frac{7}{32} \zeta_3 \xi^3
\right\}
\nn \\
&& + {\cal{O}}(\ep),
\eea
\bea
p^2 c^{(2,\xi)}(0,p^2,0) & \!\!=\!\! &
C_A^2 \; \frac{g^4\;\eta^2}{(4\pi)^n}
(-p^2)^{-2\ep}
\left\{
\frac{1}{\ep^4}
\left(  \frac{5}{32} \!-\! \frac{3}{32} \xi \right)
+ \frac{1}{\ep^3}
\left(  \frac{109}{48} \!+\! \frac{31}{96} \xi
\!-\! \frac{3}{128} \xi^2 \!-\! \frac{45}{256} \xi^3 \right) 
\right.
\nn \\
&&  + \frac{1}{\ep^2}
\left(  \frac{775}{72} \!+\! \frac{1}{32} \pi^2
   \!+\! \frac{1231}{576} \xi \!-\! \frac{5}{96} \pi^2 \xi
   \!+\! \frac{11}{32} \xi^2 \!-\! \frac{1}{192} \pi^2 \xi^2
   \!-\! \frac{1}{4} \xi^3 \!-\! \frac{1}{192} \pi^2 \xi^3 \right)
\nn \\
&& + \frac{1}{\ep}
\left( \frac{1931}{54} + \frac{19}{48} \pi^2 - \frac{3}{2} \zeta_3
+ \frac{185}{27} \xi - \frac{11}{72} \pi^2 \xi
- \frac{1}{2} \zeta_3 \xi + \frac{5}{8} \xi^2
+ \frac{7}{128} \pi^2 \xi^2  
\right.
\nn \\
&& \left.
 - \frac{13}{32} \zeta_3 \xi^2
 - \frac{29}{32} \xi^3 + \frac{1}{256} \pi^2 \xi^3  
- \frac{1}{32} \zeta_3 \xi^3 \right)
\nn \\
&&  + \frac{17443}{162} + \frac{107}{48} \pi^2
- 15 \zeta_3 - \frac{3}{160} \pi^4
+ \frac{70727}{2592} \xi - \frac{755}{1728} \pi^2 \xi
- \frac{317}{48} \zeta_3 \xi 
\nn \\   
&&  - \frac{3}{160} \pi^4 \xi + \frac{45}{16} \xi^2
+ \frac{7}{32} \pi^2 \xi^2
- \frac{57}{32} \zeta_3 \xi^2 - \frac{1}{128} \pi^4 \xi^2 
- \frac{115}{64} \xi^3 - \frac{7}{128} \pi^2 \xi^3
\nn \\
&& \left.
+ \frac{27}{64} \zeta_3 \xi^3  - \frac{1}{640} \pi^4 \xi^3
\right\}
+ {\cal{O}}(\ep),
\eea
\bea
p^2 d^{(2,\xi)}(0,p^2,0) & \!\!=\!\! &
C_A^2 \; \frac{g^4\;\eta^2}{(4\pi)^n}
(-p^2)^{-2\ep}
\left\{
\frac{1}{\ep^4}
\left( \frac{3}{16} - \frac{1}{32} \xi
- \frac{1}{128} \xi^2  - \frac{1}{128} \xi^3 \right)
\right.
\nn \\
&& + \frac{1}{\ep^3}
\left(  - \frac{1}{48} + \frac{43}{48} \xi - \frac{15}{64} \xi^2
- \frac{7}{64} \xi^3  \right)
\nn \\
&& + \frac{1}{\ep^2} 
\left( - \frac{875}{144} + \frac{13}{48} \pi^2
+ \frac{1429}{288} \xi  - \frac{5}{64} \pi^2 \xi
- \frac{25}{32} \xi^2 - \frac{85}{256} \xi^3 - \frac{1}{384} \pi^2 \xi^3
\right)
\nn \\
&& + \frac{1}{\ep}
\left( - \frac{27539}{864} + \frac{61}{48} \pi^2 + 2 \zeta_3
+ \frac{32149}{1728} \xi - \frac{17}{96} \pi^2 \xi
- \frac{21}{32} \zeta_3 \xi
- \frac{127}{32} \xi^2
\right.
\nn \\
&& \left.
+ \frac{5}{64} \pi^2 \xi^2
- \frac{21}{64} \zeta_3 \xi^2
- \frac{101}{128} \xi^3 - \frac{1}{96} \pi^2 \xi^3
+ \frac{1}{32} \zeta_3 \xi^3 \right)
\nn \\ 
&& - \frac{651695}{5184} + \frac{541}{144} \pi^2
+ \frac{67}{4} \zeta_3 + \frac{7}{80} \pi^4
+ \frac{736825}{10368} \xi - \frac{7}{6} \pi^2 \xi
- \frac{97}{16} \zeta_3 \xi 
\nn \\
&&  - \frac{17}{640} \pi^4 \xi
- \frac{841}{64} \xi^2 + \frac{43}{192} \pi^2 \xi^2
- \frac{3}{16} \zeta_3 \xi^2 - \frac{7}{1280} \pi^4 \xi^2
- \frac{17}{8} \xi^3 
\nn \\
&& \left.
- \frac{23}{768} \pi^2 \xi^3 + \frac{7}{32} \zeta_3 \xi^3
\right\}
+ {\cal{O}}(\ep),
\eea
\bea
p^2 e^{(2,\xi)}(0,p^2,0) & \!\!=\!\! &
-C_A^2 \; \frac{g^4\;\eta^2}{(4\pi)^n}
(-p^2)^{-2\ep}
\left\{  
\frac{1}{\ep^3}
\left(  \frac{1}{16} + \frac{1}{32} \xi \right)
\right.
\nn \\
&& + \frac{1}{\ep^2}
\left( \frac{23}{24}  + \frac{1}{96} \pi^2
+ \frac{3}{16} \xi - \frac{1}{192} \pi^2 \xi 
- \frac{11}{32} \xi^2 + \frac{1}{48} \pi^2 \xi^2 \right)
\nn \\
&& + \frac{1}{\ep}
\left( \frac{1201}{288} + \frac{1}{48} \pi^2 + \frac{1}{16} \zeta_3
- \frac{15}{64} \xi  - \frac{1}{32} \pi^2 \xi - \frac{1}{32} \zeta_3 \xi
- \frac{15}{32} \xi^2
\right.
\nn \\
&& \left.
+ \frac{1}{48} \pi^2 \xi^2
+ \frac{1}{8} \zeta_3 \xi^2 \right)
\nn \\
&&  + \frac{22921}{1728} + \frac{35}{144} \pi^2
- \zeta_3 + \frac{1}{320} \pi^4
- \frac{237}{128} \xi - \frac{5}{96} \pi^2 \xi
- \frac{3}{8} \zeta_3 \xi
\nn \\   
&& \left.
- \frac{1}{640} \pi^4 \xi
- 2 \xi^2 + \frac{13}{96} \pi^2 \xi^2
- \frac{1}{16} \zeta_3 \xi^2 + \frac{1}{160} \pi^4 \xi^2
\right\}
+ {\cal{O}}(\ep),
\eea
\bea
a^{(2,q)}(0,p^2,0) & \!=\! &
C_A T \; \frac{g^4\;\eta^2}{(4\pi)^n}
(-p^2)^{-2\ep}
\left\{
- \frac{1}{24\ep^3} - \frac{1}{36\ep^2} 
+ \frac{1}{\ep} \left( - \frac{23}{27} + \frac{1}{36} \pi^2 \right) 
\right. 
\nn \\
&& \left. 
- \frac{2339}{648} + \frac{5}{108} \pi^2 + \frac{5}{12} \zeta_3 
\right\}
+{\cal{O}}(\ep),
\eea
\bea
p^2 b^{(2,q)}(0,p^2,0) & \!\!=\!\! &
C_A T \; \frac{g^4\;\eta^2}{(4\pi)^n}   
(-p^2)^{-2\ep}
\left\{
\frac{1}{\ep^3}
\left( - \frac{1}{6} + \frac{1}{24} \xi \right) 
- \frac{1}{\ep^2} 
\left( \frac{25}{36} + \frac{7}{72} \xi \right) 
\right.
\nn \\
&& + \frac{1}{\ep}
\left( - \frac{1081}{216} + \frac{1}{6} \pi^2 
    - \frac{13}{27} \xi + \frac{1}{24} \pi^2 \xi \right) 
\nn \\
&& \left.
- \frac{17257}{1296} - \frac{1}{18} \pi^2 + 2 \zeta_3 
- \frac{241}{162} \xi + \frac{1}{24} \pi^2 \xi 
\right\}
+{\cal{O}}(\ep),
\eea
\bea
p^2 c^{(2,q)}(0,p^2,0) & \!\!=\!\! &
- C_A T \; \frac{g^4\;\eta^2}{(4\pi)^n}
(-p^2)^{-2\ep}
\left\{
\frac{1}{\ep^3}
\left( \frac{5}{12} + \frac{1}{12} \xi \right) 
+ \frac{1}{\ep^2} 
\left( \frac{35}{18}  + \frac{5}{36} \xi \right)
\right.
\nn \\
&& + \frac{1}{\ep}
\left( \frac{295}{54} - \frac{17}{108} \xi + \frac{1}{36} \pi^2 \xi
\right)
\nn \\
&& \left.
+ \frac{1240}{81} - \frac{5}{2} \zeta_3
- \frac{103}{81} \xi + \frac{11}{108} \pi^2 \xi - \frac{1}{3} \zeta_3 \xi
\right\}
+{\cal{O}}(\ep),
\eea
\bea
p^2 d^{(2,q)}(0,p^2,0) & \!\!=\!\! &
C_A T \; \frac{g^4\;\eta^2}{(4\pi)^n}
(-p^2)^{-2\ep}
\left\{
\frac{1}{\ep^3}
\left( \frac{5}{12} - \frac{1}{24} \xi \right) 
+ \frac{1}{\ep^2}
\left( \frac{79}{36} + \frac{7}{72} \xi \right) 
\right.
\nn \\
&& + \frac{1}{\ep}
\left( \frac{2485}{216} - \frac{1}{6} \pi^2 
    + \frac{13}{27} \xi - \frac{1}{24} \pi^2 \xi  \right) 
\nn \\
&& \left.
+ \frac{48793}{1296} + \frac{1}{9} \pi^2 - \frac{7}{2} \zeta_3
+ \frac{241}{162} \xi - \frac{1}{24} \pi^2 \xi
\right\}
+{\cal{O}}(\ep),
\eea
\be
p^2 e^{(2,q)}(0,p^2,0) =
C_A T \; \frac{g^4\;\eta^2}{(4\pi)^n}
(-p^2)^{-2\ep}
\left\{
\frac{5}{12\ep^2} + \frac{95}{72\ep} 
+ \frac{2135}{432} - \frac{1}{18} \pi^2  
\right\}   
+{\cal{O}}(\ep).
\ee

\subsection*{E.3: Non-zero out-ghost momentum squared}

\bea
a^{(2,\xi)}(0,0,p^2) & \!=\! &
C_A^2 \; \frac{g^4\;\eta^2}{(4\pi)^n}
(-p^2)^{-2\ep}
\left\{
\frac{5}{32\ep^4}
+\frac{1}{\ep^3} \left( - \frac{23}{96} + \frac{1}{16} \xi \right)
\right.
\nn \\
&& + \frac{1}{\ep^2} \left(
- \frac{143}{144} + \frac{5}{96} \pi^2 + \frac{1}{384} \pi^2\xi
+ \frac{5}{128} \xi^2 \right)
\nn \\
&& + \frac{1}{\ep} \left( - \frac{2813}{864} + \frac{1}{72} \pi^2
 + \frac{7}{8} \zeta_3 
 - \frac{33}{64} \xi  + \frac{1}{24} \pi^2 \xi
+ \frac{1}{64} \zeta_3 \xi + \frac{5}{32} \xi^2 \right)
\nn \\
&& - \frac{52535}{5184} - \frac{19}{216} \pi^2 + \frac{83}{24} \zeta_3
+ \frac{1}{40} \pi^4
- \frac{1145}{384} \xi + \frac{31}{192}  \pi^2 \xi
+ \frac{1}{2} \zeta_3 \xi 
\nn \\
&& \left.
+ \frac{1}{1280}  \pi^4 \xi
+ \frac{107}{192} \xi^2 + \frac{1}{384} \pi^2 \xi^2
- \frac{1}{32} \zeta_3 \xi^2 
\right\}
+{\cal{O}}(\ep),
\eea
\bea   
p^2 b^{(2,\xi)}(0,0,p^2) & \!\!=\!\! &
C_A^2 \; \frac{g^4\;\eta^2}{(4\pi)^n}
(-p^2)^{-2\ep}
\left\{
\frac{1}{\ep^4}
\left( \frac{7}{16} + \frac{9}{64} \xi - \frac{3}{128} \xi^2 \right) 
\right.
\nn \\
&& + \frac{1}{\ep^3} 
\left( - \frac{1}{8} - \frac{89}{192} \xi + \frac{25}{128} \xi^2
   - \frac{5}{128} \xi^3 \right)
\nn \\
&& + \frac{1}{\ep^2} 
\left(  - \frac{215}{48} \!+\! \frac{1}{8} \pi^2 
   \!-\! \frac{433}{144} \xi \!+\! \frac{1}{6} \pi^2 \xi
   \!+ \frac{105}{128} \xi^2 \!+\! \frac{1}{96} \pi^2 \xi^2
   \!-\! \frac{59}{256} \xi^3 \!+\! \frac{1}{384} \pi^2 \xi^3 \right)
\nn \\
&& + \frac{1}{\ep} 
\left( - \frac{5939}{288} + \frac{7}{144} \pi^2 
   + \frac{21}{8} \zeta_3
   - \frac{14183}{1728} \xi + \frac{53}{192} \pi^2 \xi
   + \frac{53}{32} \zeta_3 \xi
   + \frac{179}{64} \xi^2 
\right.
\nn \\
&& \left.
   + \frac{19}{384} \pi^2 \xi^2
   - \frac{11}{64} \zeta_3 \xi^2
   - \frac{113}{128} \xi^3 + \frac{1}{128} \pi^2 \xi^3  
   + \frac{1}{64} \zeta_3 \xi^3 \right)
\nn \\
&& - \frac{136931}{1728} - \frac{43}{108} \pi^2
+ \frac{367}{24} \zeta_3 + \frac{11}{160} \pi^4
- \frac{360077}{10368} \xi + \frac{55}{36} \pi^2 \xi
+ \frac{95}{16} \zeta_3 \xi 
\nn \\
&& + \frac{39}{640} \pi^4 \xi
+ \frac{255}{32} \xi^2 + \frac{35}{128} \pi^2 \xi^2
- \frac{23}{16} \zeta_3 \xi^2 - \frac{1}{1280} \pi^4 \xi^2
- 3 \xi^3 + \frac{5}{256} \pi^2 \xi^3
\nn \\
&& \left.
+ \frac{3}{16} \zeta_3 \xi^3 + \frac{1}{1280} \pi^4 \xi^3
\right\}
+ {\cal{O}}(\ep),
\eea
\bea   
p^2 c^{(2,\xi)}(0,0,p^2) & \!\!=\!\! &
C_A^2 \; \frac{g^4\;\eta^2}{(4\pi)^n}
(-p^2)^{-2\ep}
\left\{
\frac{1}{\ep^4}
\left( \frac{7}{32} + \frac{1}{8} \xi - \frac{1}{128} \xi^2
   - \frac{1}{256} \xi^3 \right)
\right.
\nn \\
&& + \frac{1}{\ep^3} 
\left( \frac{1}{16} - \frac{11}{48} \xi + \frac{11}{128} \xi^2
   - \frac{1}{256} \xi^3 \right)
\nn \\
&& + \frac{1}{\ep^2} 
\left( - \frac{41}{12} + \frac{11}{96} \pi^2
   - \frac{167}{72} \xi + \frac{17}{192} \pi^2 \xi
   + \frac{83}{128} \xi^2  + \frac{1}{192} \pi^2 \xi^2
   - \frac{7}{64} \xi^3 \right)
\nn \\
&& + \frac{1}{\ep} 
\left( - \frac{2467}{144} + \frac{25}{72} \pi^2
   + \frac{13}{8} \zeta_3
   - \frac{6961}{864} \xi + \frac{31}{144} \pi^2 \xi 
   + \frac{29}{32} \zeta_3 \xi
   + \frac{37}{16} \xi^2 
\right.
\nn \\
&& \left.
+ \frac{11}{384} \pi^2 \xi^2
   - \frac{7}{64} \zeta_3 \xi^2
    - \frac{73}{128} \xi^3 + \frac{1}{256} \pi^2 \xi^3 
   + \frac{3}{128} \zeta_3 \xi^3 \right)
\nn \\
&& - \frac{57757}{864} + \frac{161}{216} \pi^2
+ \frac{31}{3} \zeta_3 + \frac{1}{20} \pi^4
- \frac{147967}{5184} \xi + \frac{505}{864} \pi^2 \xi
+ \frac{127}{24} \zeta_3 \xi 
\nn \\
&& 
+ \frac{21}{640} \pi^4 \xi
+ \frac{231}{32} \xi^2 + \frac{17}{128} \pi^2 \xi^2
- \frac{7}{16} \zeta_3 \xi^2 - \frac{1}{1280} \pi^4 \xi^2
- \frac{133}{64} \xi^3 + \frac{1}{96} \pi^2 \xi^3
\nn \\
&&  \left.
+ \frac{3}{16} \zeta_3 \xi^3 + \frac{1}{2560} \pi^4 \xi^3
\right\}
+ {\cal{O}}(\ep),
\eea
\bea   
p^2 d^{(2,\xi)}(0,0,p^2) & \!\!=\!\! &
C_A^2 \; \frac{g^4\;\eta^2}{(4\pi)^n}
(-p^2)^{-2\ep}
\left\{
\frac{1}{\ep^4}
\left( - \frac{7}{32} - \frac{9}{64} \xi + \frac{3}{128} \xi^2
\right)
\right.
\nn \\
&& + \frac{1}{\ep^3} 
\left( - \frac{1}{48} + \frac{101}{192} \xi - \frac{25}{128} \xi^2
   + \frac{5}{128} \xi^3 \right)
\nn \\
&& + \frac{1}{\ep^2} 
\left( \frac{559}{144} \!-\! \frac{3}{32} \pi^2 
   \!+\! \frac{221}{72} \xi \!-\! \frac{31}{192} \pi^2 \xi
   \!-\! \frac{13}{16} \xi^2 \!-\! \frac{1}{96} \pi^2 \xi^2
   \!+\! \frac{59}{256} \xi^3 \!-\! \frac{1}{384} \pi^2 \xi^3\!  \right) 
\nn \\
&& + \frac{1}{\ep} 
\left( \frac{16447}{864} - \frac{29}{144} \pi^2 - \frac{3}{2} \zeta_3
   + \frac{13589}{1728} \xi - \frac{37}{192} \pi^2 \xi
   - \frac{13}{8} \zeta_3 \xi
   - \frac{179}{64} \xi^2 
\right.
\nn \\
&& \left.
   - \frac{19}{384} \pi^2 \xi^2 
   + \frac{11}{64} \zeta_3 \xi^2
   + \frac{113}{128} \xi^3 - \frac{1}{128} \pi^2 \xi^3
   - \frac{1}{64} \zeta_3 \xi^3 \right)
\nn \\
&& + \frac{393163}{5184} - \frac{77}{216} \pi^2
- \frac{251}{24} \zeta_3 - \frac{7}{160} \pi^4
+ \frac{333023}{10368} \xi - \frac{341}{288} \pi^2 \xi
- \frac{87}{16} \zeta_3 \xi 
\nn \\
&&  - \frac{19}{320} \pi^4 \xi
- \frac{257}{32} \xi^2 - \frac{13}{48} \pi^2 \xi^2
+ \frac{23}{16} \zeta_3 \xi^2 + \frac{1}{1280} \pi^4 \xi^2
+ 3 \xi^3 - \frac{5}{256} \pi^2 \xi^3
\nn \\
&& \left.
- \frac{3}{16} \zeta_3 \xi^3 - \frac{1}{1280} \pi^4 \xi^3 
\right\}
+ {\cal{O}}(\ep),
\eea
\bea   
p^2 e^{(2,\xi)}(0,0,p^2) & \!\!=\!\! &
C_A^2 \; \frac{g^4\;\eta^2}{(4\pi)^n}
(-p^2)^{-2\ep}
\left\{
\frac{1}{\ep^3}
\left( \frac{1}{8} + \frac{1}{16} \xi \right) 
\right.
\nn \\
&& -  \frac{1}{\ep^2}
\left( \frac{1}{2}  + \frac{1}{48} \pi^2
   + \frac{1}{16} \xi + \frac{1}{192} \pi^2 \xi
   - \frac{3}{32} \xi^2 - \frac{1}{384} \pi^2 \xi^2 \right) 
\nn \\
&& - \frac{1}{\ep}
\left( \frac{39}{16} \!+\! \frac{1}{12} \pi^2 
    \!+\! \frac{1}{8} \zeta_3
    \!-\! \frac{23}{32} \xi \!+\! \frac{1}{96} \pi^2 \xi
    \!+\! \frac{1}{32} \zeta_3 \xi
    \!+\! \frac{3}{32} \xi^2 \!-\! \frac{1}{96} \pi^2 \xi^2
    \!-\! \frac{1}{64} \zeta_3 \xi^2 \! \right)
\nn \\
&& - \frac{929}{96} - \frac{13}{48} \pi^2 
+ \frac{3}{4} \zeta_3 - \frac{1}{160} \pi^4
+ \frac{499}{192} \xi - \frac{1}{96} \pi^2 \xi
- \frac{1}{8} \zeta_3 \xi - \frac{1}{640} \pi^4 \xi 
\nn \\
&& \left.
- \frac{85}{96} \xi^2 + \frac{1}{32} \pi^2 \xi^2
+ \frac{9}{32} \zeta_3 \xi^2 + \frac{1}{1280} \pi^4 \xi^2 
\right\}
+{\cal{O}}(\ep),
\eea
\bea
a^{(2,q)}(0,0,p^2) & \!=\! &
C_A T \; \frac{g^4\;\eta^2}{(4\pi)^n}
(-p^2)^{-2\ep}
\left\{
\frac{1}{6\ep^3} + \frac{19}{36\ep^2} 
+ \frac{1}{\ep} \left( \frac{319}{216} + \frac{1}{72} \pi^2 \right) 
\right.
\nn \\
&& \left. 
+ \frac{5245}{1296} + \frac{23}{216} \pi^2 - \frac{11}{12} \zeta_3
\right\}
+{\cal{O}}(\ep),
\eea   
\bea
p^2 b^{(2,q)}(0,0,p^2) & \!\!=\!\! &
C_A T \; \frac{g^4\;\eta^2}{(4\pi)^n}
(-p^2)^{-2\ep}
\left\{
\frac{1}{\ep^3}
\left( \frac{1}{2} + \frac{5}{24} \xi \right) 
+ \frac{1}{\ep^2}
\left( \frac{37}{12} + \frac{25}{72} \xi \right) 
\right.
\nn \\
&& + \frac{1}{\ep}
\left( \frac{469}{72} + \frac{4}{9} \pi^2 
    + \frac{13}{27} \xi + \frac{1}{24} \pi^2 \xi \right) 
\nn \\
&& \left.
+ \frac{9085}{432} + \frac{38}{27} \pi^2 - \frac{1}{3} \zeta_3 
+ \frac{79}{162} \xi + \frac{11}{72} \pi^2 \xi - \zeta_3 \xi
\right\}
+{\cal{O}}(\ep),
\eea
\bea
p^2 c^{(2,q)}(0,0,p^2) & \!\!=\!\! &
C_A T \; \frac{g^4\;\eta^2}{(4\pi)^n}
(-p^2)^{-2\ep}
\left\{
\frac{1}{\ep^3}
\left( \frac{1}{2} + \frac{1}{12} \xi \right) 
+ \frac{1}{\ep^2}
\left(  \frac{17}{6} + \frac{2}{9} \xi \right) 
\right.
\nn \\  
&& + \frac{1}{\ep}
\left( \frac{341}{36} + \frac{5}{36} \pi^2 
   + \frac{4}{27} \xi + \frac{1}{36} \pi^2 \xi \right)
\nn \\ 
&& \left.
+ \frac{6503}{216} + \frac{79}{108} \pi^2 - \frac{13}{6} \zeta_3
- \frac{19}{81} \xi + \frac{7}{54} \pi^2 \xi 
- \frac{1}{3} \zeta_3 \xi
\right\}
+{\cal{O}}(\ep),
\eea
\bea
p^2 d^{(2,q)}(0,0,p^2) & \!\!=\!\! &
- C_A T \; \frac{g^4\;\eta^2}{(4\pi)^n}
(-p^2)^{-2\ep}
\left\{
\frac{1}{\ep^3}
\left(  \frac{1}{3} + \frac{5}{24} \xi \right) 
+ \frac{1}{\ep^2}
\left( \frac{89}{36} + \frac{25}{72} \xi \right) 
\right.
\nn \\
&& + \frac{1}{\ep}
\left(  \frac{1049}{216} + \frac{17}{36} \pi^2
    + \frac{13}{27} \xi + \frac{1}{24} \pi^2 \xi \right) 
\nn \\
&& \left.
+ \frac{22205}{1296} + \frac{163}{108} \pi^2 + \frac{5}{6} \zeta_3 
+ \frac{79}{162} \xi + \frac{11}{72} \pi^2 \xi - \zeta_3 \xi
\right\}   
+{\cal{O}}(\ep),
\eea
\be
p^2 e^{(2,q)}(0,0,p^2) =
C_A T \; \frac{g^4\;\eta^2}{(4\pi)^n}
(-p^2)^{-2\ep}
\left\{
\frac{1}{2\ep^2} + \frac{7}{4\ep} 
+ \frac{155}{24} - \frac{1}{12} \pi^2 
\right\}   
+{\cal{O}}(\ep).
\ee


\newpage

\begin{figure}[htb]
\refstepcounter{figure}
\label{fig1}
\addtocounter{figure}{-1}
\phantom{AAA}
\begin{center}
\setlength{\unitlength}{1cm}
\begin{picture}(15,27)
\put(0,4)
{\mbox{\epsfysize=25cm\epsffile{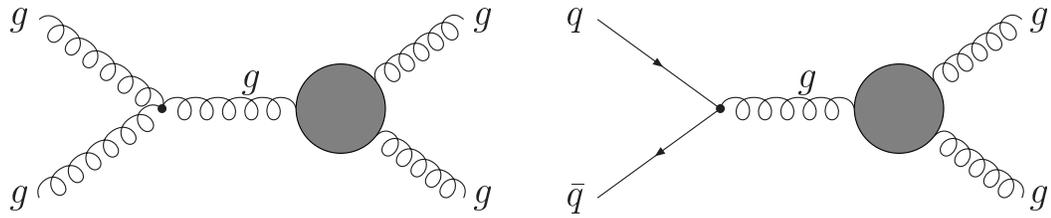}}}
\end{picture}
\end{center}
\vspace*{-22.0cm}
\caption{Contributions to the $gg\rightarrow gg$ and 
$q\bar{q}\rightarrow gg$ processes which involve the
three-gluon vertex}
\end{figure}

\begin{figure}[htb]
\refstepcounter{figure}
\label{fig2}
\addtocounter{figure}{-1}
\begin{center}
\setlength{\unitlength}{1cm}
\begin{picture}(15,15)
\put(0,-9)
{\mbox{\epsfysize=25cm\epsffile{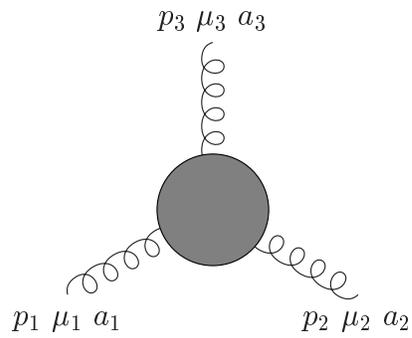}}}
\end{picture}
\end{center}
\vspace*{-8.0cm}
\caption{Notation used for gluon momenta ($p_1+p_2+p_3=0$),
Lorentz indices, $\mu_i$, and colour indices, $a_i$}
\end{figure}
\newpage

\begin{figure}[htb]
\refstepcounter{figure}
\label{fig3}
\addtocounter{figure}{-1}
\begin{center}
\setlength{\unitlength}{1cm}
\begin{picture}(15,20)
\put(-1,-5)
{\mbox{\epsfysize=25cm\epsffile{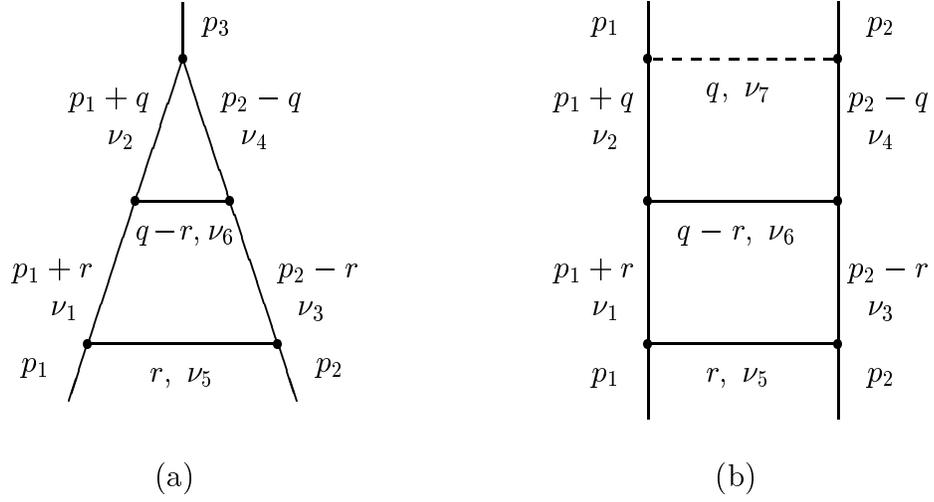}}}
\end{picture}
\end{center}
\vspace*{-10.0cm}
\caption{The planar two-loop three-point diagram (a)
        and an auxiliary four-point function (b)}
\end{figure}
\newpage

\begin{figure}[htb]
\refstepcounter{figure}
\label{fig4}
\addtocounter{figure}{-1}
\phantom{AAA}
\begin{center}
\setlength{\unitlength}{1cm}
\begin{picture}(15,27)
\put(0,4)
{\mbox{\epsfysize=25cm\epsffile{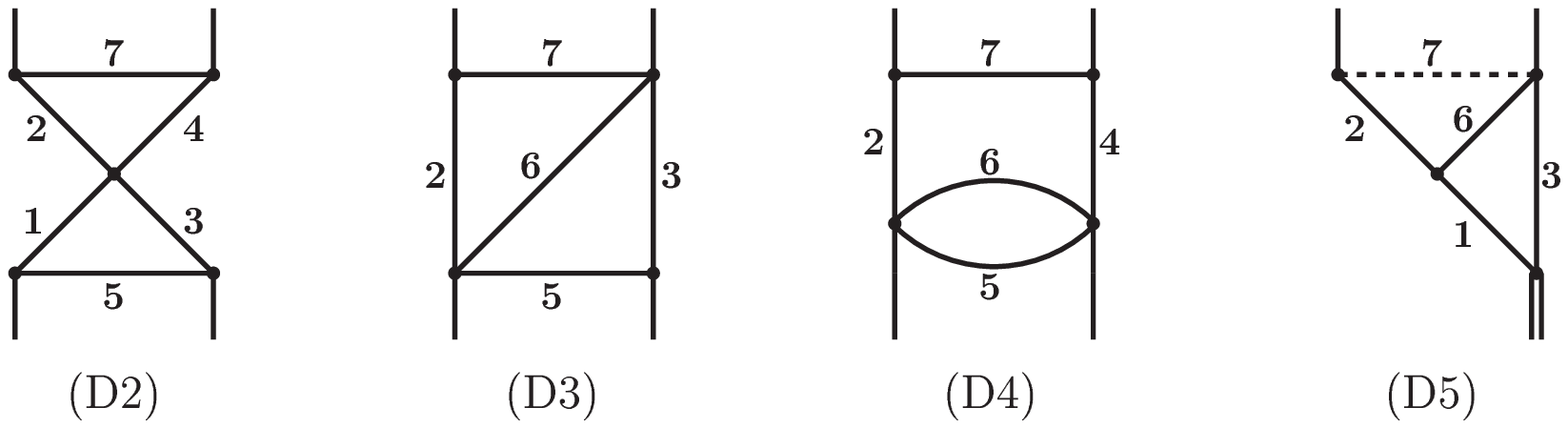}}}
\end{picture}
\end{center}
\vspace*{-21.0cm}
\caption{Boundary integrals $K_3$}
\end{figure}

\begin{figure}[htb]
\refstepcounter{figure}
\label{fig5}
\addtocounter{figure}{-1}
\begin{center}
\setlength{\unitlength}{1cm}
\begin{picture}(15,15)
\put(0,-9)
{\mbox{\epsfysize=25cm\epsffile{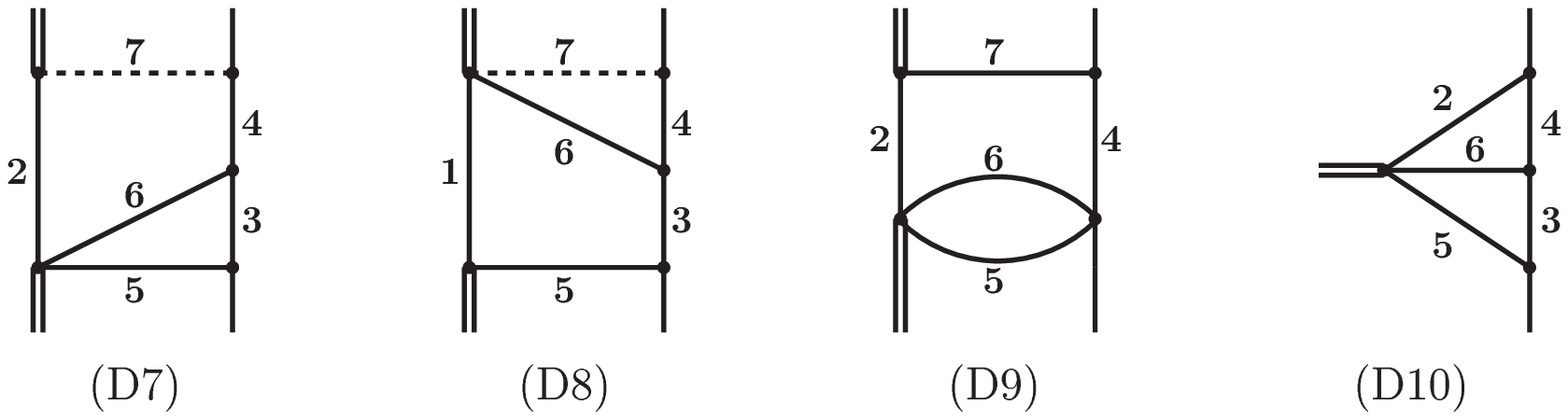}}}
\end{picture}
\end{center}
\vspace*{-8.0cm}
\caption{Boundary integrals $K_2$}
\end{figure}


\end{document}